\documentclass{aa}  

\usepackage{graphicx}
\usepackage{txfonts}
\usepackage{rotating}
\usepackage{makecell}
\usepackage[colorlinks=true]{hyperref}
\usepackage{xcolor,colortbl}
\usepackage{float}
\usepackage{adjustbox}
\usepackage{pdflscape}
\usepackage{lscape}
\usepackage{comment}
\usepackage{changepage}
\usepackage{geometry}
\usepackage{threeparttable}
\hypersetup{linkcolor=blue,citecolor=blue,urlcolor=blue}
\makeatletter
\renewcommand*\aa@pageof{, page \thepage{} of \pageref*{LastPage}}
\makeatother
\usepackage{diagbox}
\usepackage{multirow}
\usepackage{graphicx} 
\graphicspath{ {./images/} }

\usepackage[rightcaption]{sidecap}

\usepackage{wrapfig}
\usepackage{tabularx}
\usepackage[caption=false]{subfig}

\begin{document}

   \title{An automated activity classification tool for optical galaxy spectra}


   \author{C. Daoutis \inst{1,2,3}
         \and
          A. Zezas \inst{1,2,3}
          \and
          E. Kyritsis\inst{1,2}
          \and 
          K. Kouroumpatzakis \inst{4}
          \and
          P. Bonfini \inst{1,5,6}
          }

   \institute{ Physics Department, and Institute of Theoretical and Computational Physics, University of Crete, 71003 Heraklion, Greece \\
         \email{cdaoutis@physics.uoc.gr}
         \and
             Institute of Astrophysics, Foundation for Research and Technology-Hellas, 71110 Heraklion, Greece
        \and
            Center for Astrophysics | Harvard \& Smithsonian, 60 Garden St., Cambridge, MA 02138, USA
        \and 
            Astronomical Institute, Academy of Sciences, Bo\v{c}n\'{\i} II 1401, CZ-14131 Prague, Czech Republic
        \and
            ALMA Sistemi Srl, Guidonia (Rome), 00012, Italy
        \and
            Quantum Innovation Pc, Chania, 73100, Greece}

   \date{Received: 11 July 2025/ Accepted: 23 February 2026}

 
  \abstract
   {Reliable and versatile galaxy activity diagnostic tools are indispensable for comprehending the physical processes that drive galaxy evolution. Traditional methodologies frequently necessitate extensive preprocessing, such as starlight subtraction and emission line deblending (e.g., H$\alpha$ and [\ion{N}{II}]), which can introduce substantial biases and uncertainties due to their model-dependent nature. Additionally, numerous diagnostics omit the inclusion of dormant (passive) galaxies.}
   {This work aims to develop a reliable, automated, and efficient diagnostic tool capable of distinguishing between star-forming, active galactic nuclei (AGN), low-ionization nuclear emission-line regions (LINERs), composite, and passive galaxies under one unified scheme.}
   {We developed a diagnostic tool based on a support vector machine trained on ground truth data originating from optical emission-line ratios and color selection criteria. Building upon previous literature findings and exploring various combinations of discriminatory feature schemes, we identified the equivalent widths (EWs) of H$\beta$, [\ion{O}{III}]\,$\lambda$5007, and H$\alpha$+[\ion{N}{II}]\,$\lambda\lambda$6548,84 as key discriminatory features. Additionally, galaxies classified as AGN can be distinguished into broad and narrow-line AGN by measuring the full quarter at the half-maximum of H$\alpha$ and [\ion{N}{II}] complex.}
   {Employing machine learning algorithms and three EWs directly measured from the galaxy’s optical spectrum, we have developed a diagnostic tool that encompasses all potential activities of galaxies while simultaneously achieving high performance scores across all of them. Our diagnostic achieves overall accuracy of $\sim$83\% and recall of $\sim$79\% for star forming, $\sim$94\% for AGN, $\sim$85\% for LINER, $\sim$77\% for composite, and $\sim$96\% for passive galaxies.}
   {Our diagnostic tool offers significant improvements over the existing galaxy activity diagnostics as it can be applied to large numbers of spectra, eliminates the need for preprocessing (i.e., starlight subtraction or flux calibration) and spectral line deblending, encompasses all activity classes under one unified scheme, and offers the ability to distinguish between the two main types of AGN. In addition, the omission of starlight subtraction was not found to significantly reduce the diagnostic's performance. Furthermore, the narrow wavelength range required for its application enables its use to a wide range of redshifts, making it highly relevant to activity studies of high-redshift galaxies.}


   \maketitle
%

\section{Introduction}

The characterization of a galaxy’s activity is of great significance for many areas of extragalactic astrophysics, yet it often presents challenges due to the intricate interplay of several astrophysical processes involved. Galaxies exhibit a diverse range of activities, from quiescent (passive) characterized by minimal to absent star formation to active, marked by intense starburst activity or accretion onto their or supermassive black hole. Furthermore, high-coverage sky surveys, such as the Sloan Digital Sky Survey (SDSS) or The Large Sky Area Multi-Object Fiber Spectroscopic Telescope \citep[LAMOST;][]{2012RAA....12.1197C}, routinely produce substantial datasets of galaxies, providing valuable insights into galaxy populations and their properties but also posing challenges for traditional data analysis methods. This highlights the need for an automated activity classification tool with high accuracy that could be applied to large number of galaxy samples, and include a wide range of galaxy activities.

The optical part of a galaxy's spectrum holds a wealth of information pertinent to its activity. Thus, many diagnostic methods have been developed over the years trying to classify galaxy activity using emission lines such as the H$\alpha$, H$\beta$, [\ion{O}{III}] $\lambda$5007, [\ion{N}{II}] $\lambda$6584, [\ion{S}{II}] $\lambda\lambda$6717,6731, and [\ion{O}{I}] $\lambda$6300 \citep[hereafter BPT diagrams; e.g.,][]{1981PASP...93....5B,1987ApJS...63..295V,2001ApJ...556..121K,2003MNRAS.346.1055K,2007MNRAS.382.1415S} with the most widely used being the [\ion{O}{III}]/H$\beta$ versus [\ion{N}{II}]/H$\alpha$. The success of these lines can be attributed to the fact that they are sensitive to ionizing photons of different energy and hence continua of different hardness \citep{2019ARA&A..57..511K}. However, estimating the flux for an emission line requires line profile fitting, which can be model-dependent, particularly for blended lines (e.g., [\ion{N}{II}] and H$\alpha$) that may have multiple components. In addition AGN can be classified  into broad-line AGN (Seyfert 1) and narrow-line (AGN) depending on the the presence of broad permitted emission lines. In particular, true Seyfert 1 AGN (as opposed to intermediate types such as 1.9–1.1) are typically characterized by H$\alpha$ FWHM $\gtrsim$ 1000 km s$^{-1}$ \citep[e.g.,][]{2005ApJ...630..122G}, while many studies adopt a more conservative threshold of FWHM $\gtrsim$ 2000 km s$^{-1}$ \citep[e.g.,][]{1985ApJ...297..166O,2000yCat.7215....0V}. Objects with FWHM in the range $\sim$500-2000 km s$^{-1}$ are usually classified as narrow-line Seyfert 1 galaxies \citep[e.g.,][]{2006ApJS..166..128Z}. For the traditional diagnostics only the narrow-line components are used, further necessitating the need for line profile fitting.

Furthermore, the accurate flux estimation requires starlight subtraction, which can introduce biases resulting from the assumptions about the underlying stellar populations. Although line fluxes measured in starlight subtracted spectra are generally robust \citep[e.g.,][]{2014MNRAS.441.2296M}, the starlight subtraction process introduces uncertainties that are hard to quantify and in the case of spectra with strong AGN continuum may bias the results since it only accounts for the stellar component. Lastly, emission line diagnostics often exclude objects with weak or absent activity (i.e., passive galaxies) from these diagnostics. In order to remedy this limitation diagnostics utilizing the EW of the H$\alpha$ line have been proposed \citep{2010MNRAS.403.1036C}. However, these diagrams still rely on the line flux of H$\alpha$ and [\ion{N}{II}] $\lambda$6584 requiring profile fitting and starlight subtraction.

In this work, motivated by the results presented in \cite{2025A&A...693A..95D}, we propose an automated diagnostic tool for galaxy activity. In contrast to the aforementioned work, our approach focuses on automating the activity characterization of galaxies by using spectra without the need for starlight subtraction. In addition, while that work demonstrated remarkable success in the identification and decomposition of galaxy activity, it uses the D4000 index break \citep{1999ApJ...527...54B}. This may hinder its application to low-quality spectra, as the blue portion of the spectrum can be challenging to measure reliably. Furthermore, in our diagnostic we include all classes including broad-line AGN in addition to the narrow line ones considered in the traditional diagnostics. Specifically, we encompass five distinct activity classes: star-forming, AGN, LINER, composite, and passive galaxies and utilize three features: the EWs of [\ion{O}{III}], H$\beta$, and the combined H$\alpha$ + [\ion{N}{II}] $\lambda \lambda$6548,84. These features are measured directly on the observed spectra, eliminating the need for starlight subtraction and deblending of the [\ion{N}{II}] doublet and the H$\alpha$ line. Another advantage of this method is that since it is based on the EW of all the considered lines it is not sensitive to extinction effects and does not require flux calibration, making it particularly useful for multi-fiber spectra. 

This paper is organized as follows. In Section \ref{Data_sample}, we introduce the galaxy sample and the methods employed to obtain the ground truth labels for the activity classification of the various classes. In Section \ref{Developing_the_diagnostic_tool}, the development of the diagnostic is described in detail, as well as the process for identifying broad- and narrow-line AGN. Additionally, we present the algorithm employed and the metrics utilized to assess its performance. In Section \ref{Results}, we present the results and the diagnostic’s performance for classifying and identifying the AGN subclasses. In Section \ref{Disussion}, we present the outcomes of a diagnostic applied to spectra of different spectral resolution or signal-to-noise ratio. We discuss methods for assessing the classification reliability and the influence of the stellar continuum on the classification of and AGNs, particularly in the case of low AGN fraction contribution. Finally, the overall conclusions are presented in Section \ref{Conclusions}.

\section{Data sample} \label{Data_sample}

\subsection{The galaxy sample}

Our galaxy sample is primarily drawn from the SDSS survey. The Sloan Digital Sky Survey \citep[SDSS;][]{2000AJ....120.1579Y} has revolutionized the study of galaxies by providing high-quality optical spectra for millions of galaxies across a wide range of environments and redshifts. SDSS spectra are uniformly processed, ensuring consistency across the dataset, and they cover a large portion of the sky, making them ideal for statistical studies of galaxy evolution. They cover the 3800-9200 Å spectral range, encompassing all the diagnostically important optical lines with a resolving power ranging from \( R \sim 1500 \) at 3800 Å to \( R \sim 2500 \) at 9000 Å for the original SDSS-I/II/III spectrographs. The Baryon Oscillation Spectroscopic Survey (BOSS) and the extended BOSS (eBOSS) used in SDSS-IV, extends the wavelength coverage to 3600–10400 Å with a resolution of \( R \sim 1300 \) at 3600 Å and \( R \sim 2500 \) at 10400 Å. We used the same galaxy sample introduced in \cite{2025A&A...693A..95D} which was built by combining spectroscopic data from the MPA-JHU DR8 \citep{2003MNRAS.346.1055K,2004MNRAS.351.1151B,2004ApJ...613..898T} release of the SDSS and ultraviolet photometry from the GALEX survey, via the GSWLC catalog \citep{2016ApJS..227....2S}. Since our analysis in this project differs from that of \cite{2025A&A...693A..95D}, we applied small modifications to these criteria to tailor the sample to better the needs of this project.

\subsection{Galaxy activity classification} \label{gal_act_classes}

In order to define our new diagnostic we need a sample with reliable activity classifications for all the activity types we consider in this project. These will be used to delineate the locus of each activity class in the parameter space defined by the spectral features we consider in our diagnostic.

Galaxies that show prominent emission lines, namely star-forming, AGN, LINER, and composite, are selected by implementing the Soft Data-Driven Allocation \cite[SoDDA;][]{2019MNRAS.485.1085S} classifier, a four-dimensional diagnostic based jointly on the emission-line ratios of log$_{10}$([\ion{N}{II}]/H$\alpha$), log$_{10}$([\ion{S}{II}]/H$\alpha$), log$_{10}$([\ion{O}{I}]/H$\alpha$), and log$_{10}$([\ion{O}{III}]/H$\beta$). This method presents a significant advantage in comparison to the commonly used two dimensional emission-line ratio diagnostics since it simultaneously considers all four crucial features preventing contradictory classifications. The emission line fluxes were acquired from the MPA-JHU catalog which provides deblended, starlight subtracted line measurements for SDSS spectra obtained up to the SDSS DR7. 

The sample of passive galaxies was defined using the $NUV - r$ against the absolute magnitude in the SDSS $r$-band ($M_{r}$) color-magnitude diagram which it was found to be the most reliable criterion with almost no contamination from green valley galaxies \citep{2008MNRAS.385.1201H}.

\subsection{Characterization of AGN subclasses} \label{def_AGN_subclasses}

Following our activity characterization of galaxies, we partitioned the AGN identified with the SoDDA diagnostic (Sect. \ref{gal_act_classes}) into two subclasses: broad-line (Type 1; BL AGN) and narrow-line (Type 2; NL AGN) AGNs based on their SDSS classification. We characterize as BL AGN any galaxy classified as AGN by SoDDA and has ("\texttt{SUBCLASS}"="\texttt{BROADLINE}" or "\texttt{AGN BROADLINE}") designation from SDSS. These are objects of non stellar spectra with emission lines detected at the 10$\sigma$ level which have velocity dispersions of the Balmer emission lines exceeding 200 km s$^{-1}$ at the 5$\sigma$ level (or FWHM of $\sim$ 470 km s$^{-1}$). 

\subsection{Final sample} \label{final_sample}

For all active galaxies, we adopt the quality criteria imposed by \cite{2025A&A...693A..95D} based on the signal-to-noise ratio (S/N > 5) of the emission lines of [\ion{O}{III}], [\ion{N}{II}], [\ion{O}{I}], [\ion{S}{II}], H$\alpha$, and H$\beta$. This ensures the quality of the classification labels based on the SoDDA diagnostic. In addition, we remove any object where the MPA-JHU pipeline failed to provide reliable line measurements (\texttt{RELIABLE}=0 flag).

For the passive galaxies, all S/N based quality criteria imposed by \cite{2025A&A...693A..95D} are retained. However, we eliminate from the sample of passive galaxies any galaxy that is classified as star-forming, AGN, LINER or composite. This eliminates from our sample of passive galaxies any galaxy in which any of the diagnostic emission lines have S/N$>$5. Galaxies that exhibit emission lines at lower confidence are included in our sample in order not to be biased against excitation by old stellar populations or very weak AGN.

For all galaxies that satisfy the aforementioned criteria, we download their optical spectra from the DR8 {SDSS}\footnote[1]{\url{https://www.sdss.org/}}. The final sample composition per class is presented in Table \ref{table:sample_comp}.

\begin{table}[h]
\caption{The composition of the final sample for each galaxy class.} 
\centering 
\begin{tabular}{l c c}
\hline\hline
  Class & Number of objects & Percentage (\%)\\
\hline
Star-forming & 38381 & 61.6 \\
AGN (BL) & 235 & \multirow{2}{*}{3.6} \\  
AGN (NL) & 2016 &  \\
LINER & 1105 & 1.7 \\                     
Composite & 2610 & 4.2 \\
Passive & 18072 & 28.9 \\ 
\hline
Total & 62419 & 100 \\
\hline
\end{tabular}
\label{table:sample_comp} 
\end{table}

\subsection{Feature selection} \label{feat_selection}

Following the results presented in \cite{2025A&A...693A..95D} and motivated by previous works using the EW of H$\alpha$ as a diagnostic feature \citep{2010MNRAS.403.1036C}, we adopt the EWs of the spectral lines of H$\beta$, [\ion{O}{iii}], and H$\alpha$ + [\ion{N}{ii}] as diagnostic indicators. Utilizing EWs instead of spectral line fluxes offers two key advantages: it allows for the inclusion of galaxies with very faint or nonexistent emission lines, e.g., passive galaxies, and it is reddening insensitive.

While the labels for the spectral classification of the emission line galaxies are based on the starlight-subtracted line measurements and the SoDDA diagnostic, in our diagnostic we measured the EWs directly from the reduced SDSS spectra processed by the SDSS pipeline without performing any starlight subtraction. This is because our goal is to develop a diagnostic applicable to spectra requiring minimal processing. We adhere to the convention established by the SDSS, which assigns negative EW values to emission lines. Furthermore, we measure the combined EW of the [\ion{N}{ii}] doublet and H$\alpha$ which eliminates the need for profile fitting and deblending which can be model depended, especially in objects with broad-line components. In summary, our feature scheme consist of the EWs of H$\beta$, [\ion{O}{iii}] $\lambda$5007, and the H$\alpha$ + [\ion{N}{II}] $\lambda \lambda$6548,84. 

\begin{table*}[ht]
\caption{The spectral features, central wavelengths, wavelength ranges, and continuum ranges used for the calculation of the EWs.}
\centering
\begin{tabular}{l c c c c c}
\hline\hline
Spectral feature & Center (\text{$\AA$}) & Wavelength span (\text{$\AA$}) & Continuum left (\text{$\AA$}) & Continuum right (\text{$\AA$}) \\
\hline
H$\beta$ & 4864 & 30 & 4839-4849 & 4879-4889 \\
$[\ion{O}{III}]$ $\lambda 5007$ & 5007 & 30 & 4982-4992 & 5022-5032 \\
H$\alpha$ + [\ion{N}{II}] $\lambda \lambda$6548,84 & 6566 & 80 & 6516-6526 & 6606-6616 \\
\hline
\end{tabular}
\label{EW_bands}
\end{table*}

The wavelength ranges of these spectral bands are defined around the central wavelength of the relevant spectral lines. Their wavelength ranges are 80 \text{$\AA$} for the H$\alpha$ + [\ion{N}{II}] blend, and 30 \text{$\AA$} for the [\ion{O}{III}] and H$\beta$. In Table \ref{EW_bands} we report the centers of each targeted feature, their wavelength span, and the range of the continuum on either side of each spectral window. The estimation of the continuum is performed from a 10 \text{$\AA$} wide region on either side of each spectral line, using all points and their corresponding uncertainties, and fitting linear model using the weighted least squares method. This model is then used for the estimation of the continuum at each wavelength point in each spectral line band of interest for the calculation of its EW (see Appendix \ref{AppC}, equation \ref{eq:equivalent_width}). A linear model provides a good approximation to the local continuum due to the short wavelength range and the absence of significant changes in the continuum shape. These bandwidths are sufficiently broad to encompass the full profile of each emission line, even in low-resolution spectra, yet narrow enough to ensure that the linear approximation of the continuum remains valid within the EW measurement range and to avoid contamination from neighboring spectral lines (e.g., confusion between [\ion{O}{III}] $\lambda$4959 and [\ion{O}{III}] $\lambda$5007). At the resolution of SDSS, the line band is narrow enough that the adjacent continuum region remains uncontaminated by the emission line itself, provided the line has a FWHM up to $\sim$10 $\AA$ (less than 0.1\% leakage of line flux to the continuum), corresponding to velocities of up to $\sim$700 km s$^{-1}$ for the H$\beta$ ($\sim$900 km s$^{-1}$ for H$\alpha$ and [\ion{N}{ii}]). For slightly higher contamination levels of 5\% these velocities are twice as large.

\begin{figure}[h]
  \centering
  \includegraphics[scale=0.5]{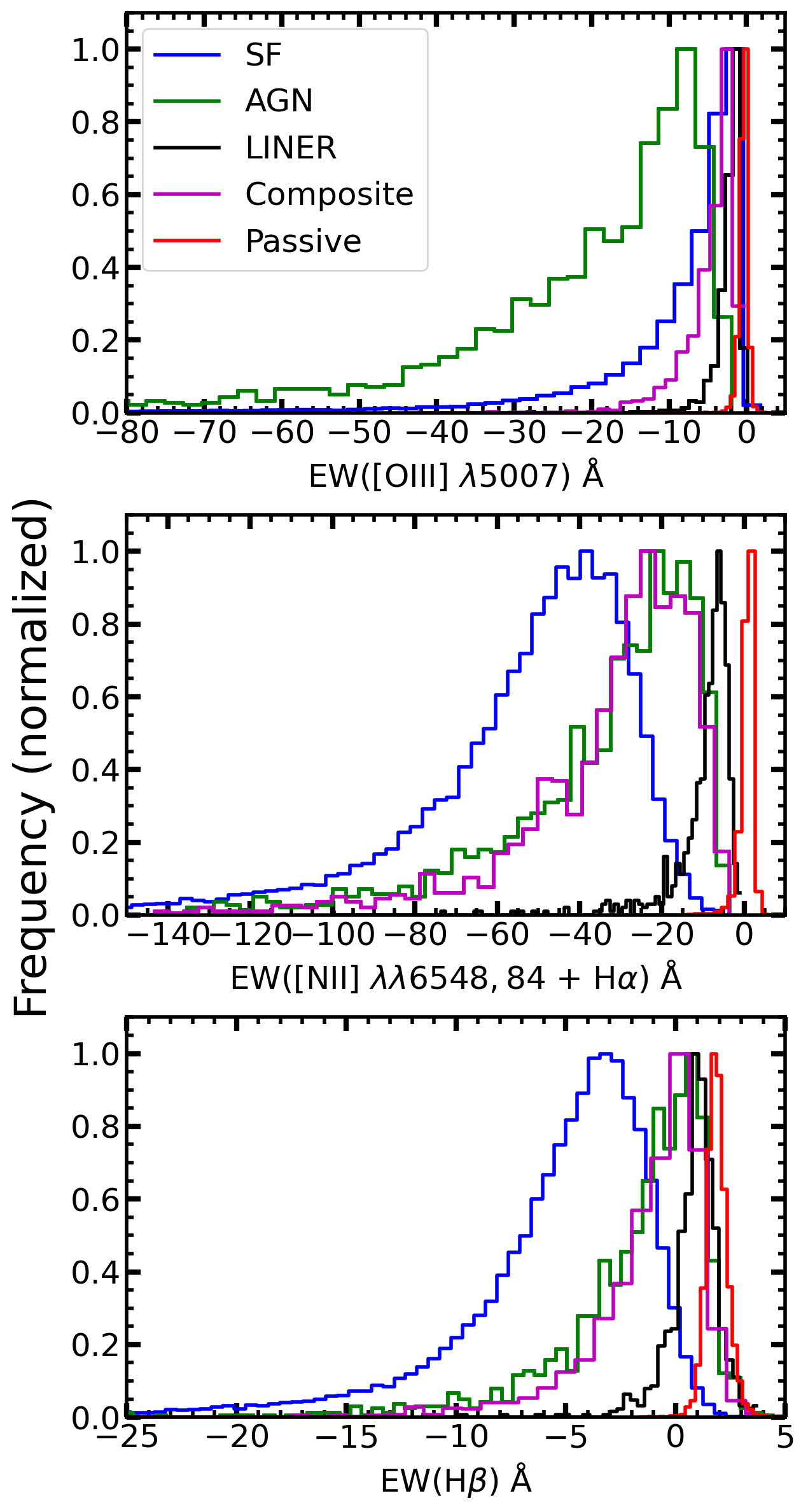}
  \caption{Distributions of the EW of the three spectral lines [\ion{O}{iii}] (top), of [\ion{N}{ii}] doublet and H$\alpha$ (middle), and H$\beta$ (bottom) used as discriminating features for the development of our new diagnostic for the activity classes of star-forming (SF), AGN, LINER, composite, and passive galaxies. All measurements were performed on non-starlight-subtracted SDSS spectra. We adhere to the same conventions as the SDSS, wherein negative EW correspond to emission.}
  \label{ft_dist}
\end{figure}

\section{Developing the diagnostic tool} \label{Developing_the_diagnostic_tool}

\subsection{The algorithm}

Machine learning algorithms are highly effective tools in identifying intricate relationships and patterns within data sets that are challenging to discern using conventional methods. Their application spans numerous scientific fields, tackling a diverse array of challenges. For example, in the field of astrophysics, these algorithms have been employed for several purposes. They have been utilized in tasks like stellar classification \citep[e.g.,][]{2022A&A...657A..62K,2022A&A...666A.122M}. Similarly, they have been employed for galaxy classification utilizing the BPT diagram \citep[e.g.,][]{2019MNRAS.485.1085S}. 

For this work we adopted the support vector machine \citep[SVM;][]{cortes1995support} algorithm. SVMs are versatile and powerful tools for classification, regression, and outlier detection, especially effective in high-dimensional spaces. They can handle non-linear relationships through kernels while offering flexibility, and efficiency making them suitable for a wide range of applications. 

SVMs function by identifying the optimal hyperplane that separates the locus of the different classes in the considered multi-dimensional space. They achieve this by focusing on the data points closest to the hyperplane, known as support vectors, which uniquely determine the separation boundary. For non-linearly separable classes, SVMs employ kernel functions to transform the data into a higher-dimensional space where a linear separation is possible, making them extremely efficient. Preliminary testing showed that, for our data, these properties made it more robust than the other alternatives we explored, namely random forest and its variants, leading us to adopt SVM as the algorithm of choice.

\subsection{Implementation} \label{implemantation}

We employ the Python 3, \texttt{scikit-learn} \citep{scikit-learn} implementation of the SVM algorithm, and specifically the Support Vector Classification (SVC, i.e., SVM for classification tasks). For the training process, the algorithm is provided with three discriminatory features: the EWs of H$\beta$, [\ion{O}{III}] $\lambda$5007, and H$\alpha$ + [\ion{N}{II}] $\lambda \lambda$6548,84 (see Sect. \ref{feat_selection}), along with the activity class of each object considered, which serves as the ground truth. In total, we consider five activity classes: star-forming (SF), AGN, LINER, composite, and passive.

As the development of a robust classifier requires high-quality data, we utilized the final sample described in Sect. \ref{final_sample}. Subsequently, these data are partitioned into two subsamples, one of which serves as the training set, while the other constitutes the test set in a 70-30\% ratio, respectively. The training set was exclusively used for the training process, while the test set was reserved for evaluating the performance of the algorithm. 

Our training dataset comprises classes that are represented by varying numbers of objects, resulting in a significant imbalance. This can lead to biases toward the majority class, thereby impeding the diagnostic’s performance for the under-represented classes. We enforced balancing by assigning class-specific weights to the penalty term within the optimization function, namely through the \texttt{class\_weight} parameter. This adjustment increases the cost associated with misclassifying samples from the minority class, thereby promoting the definition of more balanced decision boundaries.

The discriminatory features were standardized prior to their use for in training the algorithm. Standardization of input features is crucial for training an SVM, as the algorithm is sensitive to the scale of the features as it relies on distance-based computations. In other words, features with larger ranges can disproportionately influence the model, resulting in biased decision boundaries. In this study, the \texttt{StandardScaler()} from \texttt{scikit-learn} was used as the normalization method to transform each feature, ensuring that it had a mean of 0 and a standard deviation of 1. Furthermore, the SVM hyperparameters have been optimized through a random search, as described in Appendix \ref{AppA}.

\subsection{Performance metrics} \label{per_mat}

The confusion matrix serves as the main tool for assessing the performance of classification models. It presents the correctly classified objects (primary diagonal elements) and the misclassified instances (off-diagonal elements), simultaneously providing insights into the class distributions. From this matrix we derived the following metrics, used to evaluate the efficacy and performance of our diagnostic tool. Accuracy measures the proportion of correct classifications across all instances, although it may be misleading in imbalanced datasets such as ours. Recall, or sensitivity, focuses on the proportion of actual positives correctly identified by the model. The F$_{1}$ score is the harmonic mean of precision and recall, providing a balance between the two metrics. For the definitions and equations of these metrics, refer to Table 2 in \cite{2023A&A...679A..76D}.

\subsection{Predicted probabilities} \label{pred_prob}

Knowing the probability for an object to belong in each of the considered classes offers a more insightful perspective on the model’s decision-making process and it allows the assessment of the classification robustness.

SVMs are not inherently probabilistic models. Their primary objective is to identify a hyperplane that maximizes the boundary between classes, rather than estimating class probabilities. However, the SVM algorithm provided by \texttt{scikit-learn} includes a probability estimation through Platt scaling \citep{plattscaling}. This approach entails fitting a logistic regression model to the decision function scores. By mapping the distance from the decision boundary to probabilities, we can obtain class likelihoods. In general this is done by looking at the density of the objects according to their (true) type in the multiclass feature space considered. 

\begin{figure}[h]
  \centering
  \includegraphics[scale=0.55]{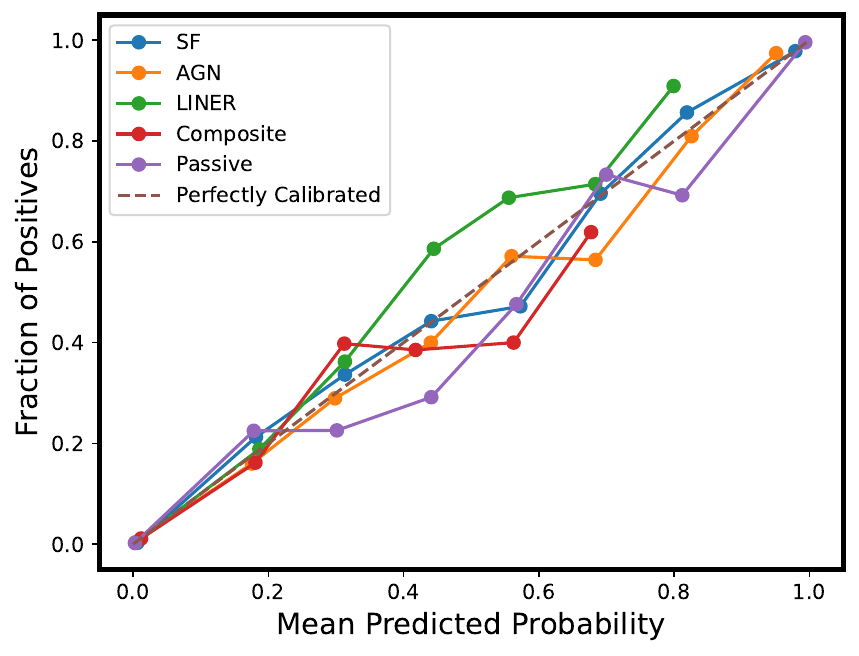}
  \caption{Calibration curves for the predicted probabilities of each activity class. This plot illustrates the relationship between the predicted probabilities (derived from our diagnostic) and the actual frequency of an activity class appearing among the remaining classes in the feature space. The dashed line represents an idealized classifier with perfect calibration. We observe that for star-forming (SF) and AGN galaxies, the predicted probabilities closely align with the observed frequencies. Passive galaxies exhibit a greater deviation from the dashed line compared to the previous two classes. Notably, LINER galaxies and passive galaxies demonstrate more pronounced deviations, which is consistent with their intricate nature.}
  \label{cal_curve}
\end{figure}

This is however not enough, since those probabilities are model-dependent, and not strictly related to the sample's demographics. A correction to this can be performed via the calibration curves, which assess the reliability of the predicted probabilities by comparing them to the corresponding true observed frequencies. To construct a calibration curve, the predicted probabilities are partitioned into bins, usually employing equal widths (i.e., equal number of objects per bin). Then, for each bin, the mean predicted probability is computed along with the actual proportion of positive instances (observed frequency). These values are subsequently plotted against each other, with the x-axis representing the mean predicted probabilities and the y-axis indicating the corresponding observed frequencies. A perfectly calibrated model will generate a diagonal line, where predicted probabilities align with observed frequencies. Deviations from this line indicate miscalibration, such as overconfidence or underconfidence in the predictions. If the calibration curves indicate a miscalibrated classifier - i.e., predicted probabilities deviate from true outcome frequencies - then post-hoc calibration techniques such as Platt scaling or isotonic regression should be applied on a separate validation set to improve probabilistic reliability. This step ensures that the classifier’s output probabilities can be meaningfully interpreted as confidence estimates. Figure \ref{cal_curve} presents the calculated calibration curves for each class implemented on our diagnostic tool. These curves indicate that our diagnostic outputs predict probabilities that closely align with the actual frequency of appearance of each class in the feature space for almost all classes, with minimal deviations, especially for AGN and star-forming galaxies. The only deviations observed appear to be limited to the classes of LINERs and composites, but only within a narrow range at the midpoint of the predicted probabilities which can be attributed to the inherently complex nature of these mixed-activity classes \citep{2023A&A...679A..76D,2025A&A...693A..95D}. This analysis suggests that the output probabilities of our diagnostic are calibrated, ensuring that the model outputs probabilities that are well-aligned with the true frequency of classes in the feature space. This makes it suitable for applications where confidence estimates or probabilistic thresholds are required, as we are interested here.

\subsection{Discriminating broad and narrow-line AGNs} \label{FWQM_BLR}

After defining our primary tool for the classification of galaxies into the five activity classes, we explored its extension to indicatively separate broad-line and narrow-line AGN. We achieved this by employing a two-step process. The first step is the five-class classifier described in the previous sections. This is followed by a separate classifier applied to the objects classified as AGN by the first step. This way the primary classifier can be readily applied even to low-quality spectra where accurately measuring the width of spectral lines may be challenging. The classifier for the broad-line and narrow-line AGN, which is more sensitive to the spectral resolution and the S/N, will not interfere with the main activity classification results. 

The secondary classifier uses only one discriminatory feature to provide information about the width of the H$\alpha$ emission line. For this discriminating feature we used the full width at the quarter maximum (FWQM) of the H$\alpha$ + [\ion{N}{II}] blend, which is the distance (in $\AA$) between two points on either side of the H$\alpha$ + [\ion{N}{II}] where the flux is found to be 1/4 of its maximum. The FWQM offered better performance than the FWHM since it probes better the wings of the line instead of its core. For the training of the secondary classifier we collected all AGN galaxies classified as broad-line and narrow line as described in Sect. \ref{def_AGN_subclasses}. The algorithm employed, the training, and the performance evaluation procedures were identical to these followed for the definition of the principal diagnostic tool (see Sect. \ref{Developing_the_diagnostic_tool}). Furthermore, to mitigate any potential biases, we utilized exactly the same AGN from the training sample (see Sect. \ref{implemantation}) of the principal classifier to train the secondary classifier. In other words, the ground truth labels used to train and test the performance of the secondary classifier are derived from dividing the AGNs in the training and test sets of the primary classifier respectively, into broad and narrow-line AGNs.

\section{Results} \label{Results}

\subsection{Performance evaluation} \label{perf_eval}

To evaluate the performance of our model, we employed the k-fold cross-validation method. This technique involves partitioning the entire dataset into k distinct subsamples (in our case, we chose k=10). In each iteration, k-1 subsamples (folds) are utilized for training, while the remaining fold serves as the evaluation set. This process was repeated k times, ensuring that each fold acts as the test set on an equal basis. By employing this approach, we obtained a reliable assessment of the model’s accuracy. This yielded a mean accuracy
across the folds of 83±1\%, indicating a high level of accuracy and robustness of our classifier.

\begin{figure}[h]
  \centering
  \includegraphics[scale=0.3]{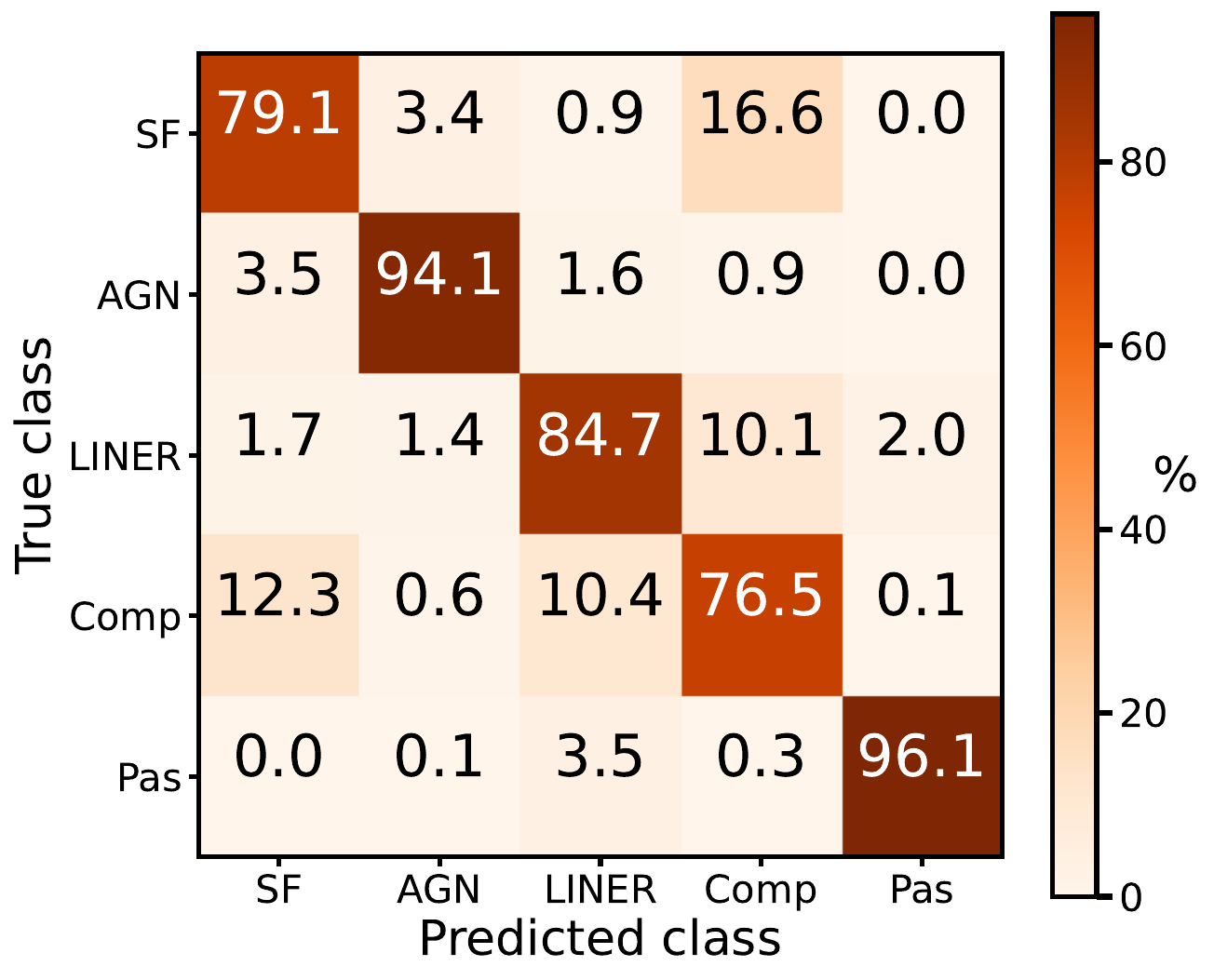}
  \caption{Confusion matrix summarizing the performance of our new diagnostic tool on the test set (Sect. \ref{implemantation}) of our final sample (Sect. \ref{final_sample}). We see that almost all objects are found on the primary diagonal indicative of a highly-performing classifier. There are a few objects in the off-diagonal elements (missclassifications), indicating mild mixing between the composite and star-forming (SF) and composite and LINER classes. This is expected, as these classes share common characteristics.}
  \label{conf_marix_principal}
\end{figure}

Table \ref{tab:p_scores} presents the performance scores for each activity class, indicating near-perfect scores throughout. The high recall values for all classes demonstrate the classifier’s ability to accurately identify almost all objects within each category, underscoring its exceptional completeness. Specifically, we observe that AGN and passive galaxies have almost perfect recall scores, with the star-forming and LINER galaxies following with very good scores. Composite galaxies, on the other hand, exhibit lower recall scores compared to the other classes, which is generally anticipated due to their nature, sharing similarities with an AGN \citep{1983ApJ...264..105F,1983ApJ...269L..37H,1984A&A...135..341S} and star-forming galaxies. Their spectra are also consistent with excitation from old stellar populations sharing some similarities with LINERs \citep{1994A&A...292...13B,2008MNRAS.391L..29S,2013A&A...555L...1P}. Indeed, examining the confusion matrix (Fig. \ref{conf_marix_principal}), we see that the primary mixing occurs between star-forming and composite galaxies. Notably, 12.3\% of composite galaxies were misclassified as star-forming, while 16.6\% of star-forming galaxies were erroneously identified as composite. Furthermore, we find that mixing between composite and LINER galaxies resulted in 10.4\% of composite galaxies being misclassified as LINERs, while 10.1\% of LINERs were incorrectly identified as composite galaxies. This behavior is anticipated, as it is widely recognized that composite and LINER galaxies can be powered by old stellar populations resulting in mixing between these populations \citep{2008MNRAS.391L..29S,2010MNRAS.403.1036C,2013A&A...558A..43S,2025A&A...693A..95D} or shocks which produce LINER-like line ratios \citep{1980A&A....87..152H,1995ApJ...455..468D,2020ApJ...893....1B,2025A&A...693A..95D}

\begin{table}[h]
\caption{Performance scores derived from the test sample for each galaxy class, utilizing three different metrics (see Sect. \ref{per_mat}).}
\centering
\begin{tabular}{l c c c c }
\hline\hline
Class & Precision & Recall & F$_{1}$-score & Number of \\
      &   &        &              &   galaxies\\ 
\hline
Star-forming & 0.99 & 0.79 & 0.88 & 11468 \\
AGN          & 0.62 & 0.94 & 0.75 & 690 \\
LINER        & 0.43 & 0.85 & 0.57 & 347 \\
Composite    & 0.23 & 0.77 & 0.36 & 787 \\
Passive      & 1.00 & 0.96 & 0.98 & 5434 \\
\hline
\end{tabular}
\label{tab:p_scores}
\end{table}

\subsection{Performance on broad and narrow-line AGN} \label{perf_on_BLR}

To evaluate the efficiency of the secondary scheme for the classification of broad and narrow-line AGN we used the same metrics as presented in Sect. \ref{per_mat}. We find that this scheme is effective in the two subclasses of AGNs, with recall rates for broad and narrow-line AGNs of 91.0\% and 90.0\%, respectively (Table \ref{table:BLR_NLR_conf_matrix}). The primary diagonal in this Table represent the recall rate, while the off-diagonal elements of this table correspond to the misclassified objects. The number of objects is indicated in parentheses. The k-fold cross-validation method is not applicable because the total number of objects is not sufficient to derive reliable statistics.

\begin{table}[h]
    \centering
    \caption{Confusion matrix for broad- and narrow-line AGN classification.}
    \begin{tabular}{lcc}
        \hline
        \diagbox{True}{Predicted} & NL-AGN  & BL-AGN  \\
        \hline
        NL-AGN  & 90.0\% (551)  & 10.0\% (61)  \\
        BL-AGN  & 9.0\%  (7)  & 91.0\% (71)  \\
        \hline
    \end{tabular}
    \tablefoot{In this classification scheme, only galaxies that have been identified as AGN by our diagnostic criteria can be subsequently categorized as broad- or narrow-line objects. Percentages reflect the total number of objects based on the ground truth (True) labels.}
    \label{table:BLR_NLR_conf_matrix}
\end{table}

The results presented in Table \ref{table:BLR_NLR_conf_matrix} demonstrate the nearly perfect efficiency of the secondary classification tool in distinguishing between broad-line and narrow-line AGNs. To further validate the above results, we visually inspected the spectra of galaxies classified by our diagnostic as broad-line and narrow-line AGNs. Our classification aligns with the spectral characteristics observed for both broad and narrow AGNs.

\section{Discussion} \label{Disussion}

In the previous sections we presented a new method for activity classification of galaxies based on the EW of the H$\beta$, [\ion{O}{iii}] $\lambda$5007, and the H$\alpha$ + [\ion{N}{II}] $\lambda \lambda$6548,84 lines. This method can characterize active galaxies of different types as well as passive galaxies, while it does not require starlight subtraction or even calibrated spectra facilitating its application to large datasets. 

\subsection{Classification confidence} \label{classification_confidence}

The confidence of the classification results depends on the intrinsic precision of the classifier (discussed in Sect. \ref{perf_eval}), on the uncertainty of the used data (in our case the EW of the diagnostic lines), and the the intrinsic mixing of the considered classes \citep{2023A&A...679A..76D}.

In order to asses the sensitivity of the classification on the measurement uncertainties affecting the EWs, we repeat the procedure described in Sect. \ref{feat_selection} on multiple realization of the spectral lines of interest, based on modeling. In particular, we model each spectral point of the spectral line and adjacent continuum as a Gaussian distribution, with the measured flux as the mean and its 1$\sigma$ uncertainty (provided by the SDSS) as the standard deviation. Utilizing this methodology, we generated a number of spectra (including the spectral line and adjacent continuum regions) from which we measured the EWs of each feature. Subsequently, we applied our classification method on each of these sets of EW measurements for each object. This process generates a distribution of classifications. Objects with low uncertainty in their spectra are anticipated to be consistently classified as belonging to a specific class. Conversely, spectra with poor quality are expected to exhibit a wide range of EW measurements, leading to frequent changes in classification across different runs, especially if they are located close to the boundary of the loci of the different classes.

An alternative method to assess the reliability of the classification is to examine the probability for an object to belong to each of the considered classes (Sect. \ref{pred_prob}). In order to account for the uncertainty of this probability stemming from the measurement uncertainties we examine the distribution of the predicted probabilities obtained from multiple runs of the diagnostic on simulated spectra, as described in the previous paragraph. In practice we calculate the mean and the standard deviation of the predicted probabilities for each source to belong to each of the considered classes. Consequently, objects with confident classifications will exhibit high probability of belonging to a single class, while probabilities  for the other classes will be low. Additionally, the standard deviation of the predicted probabilities will be low.

\begin{figure}[h]
  \resizebox{\hsize}{!}{\includegraphics{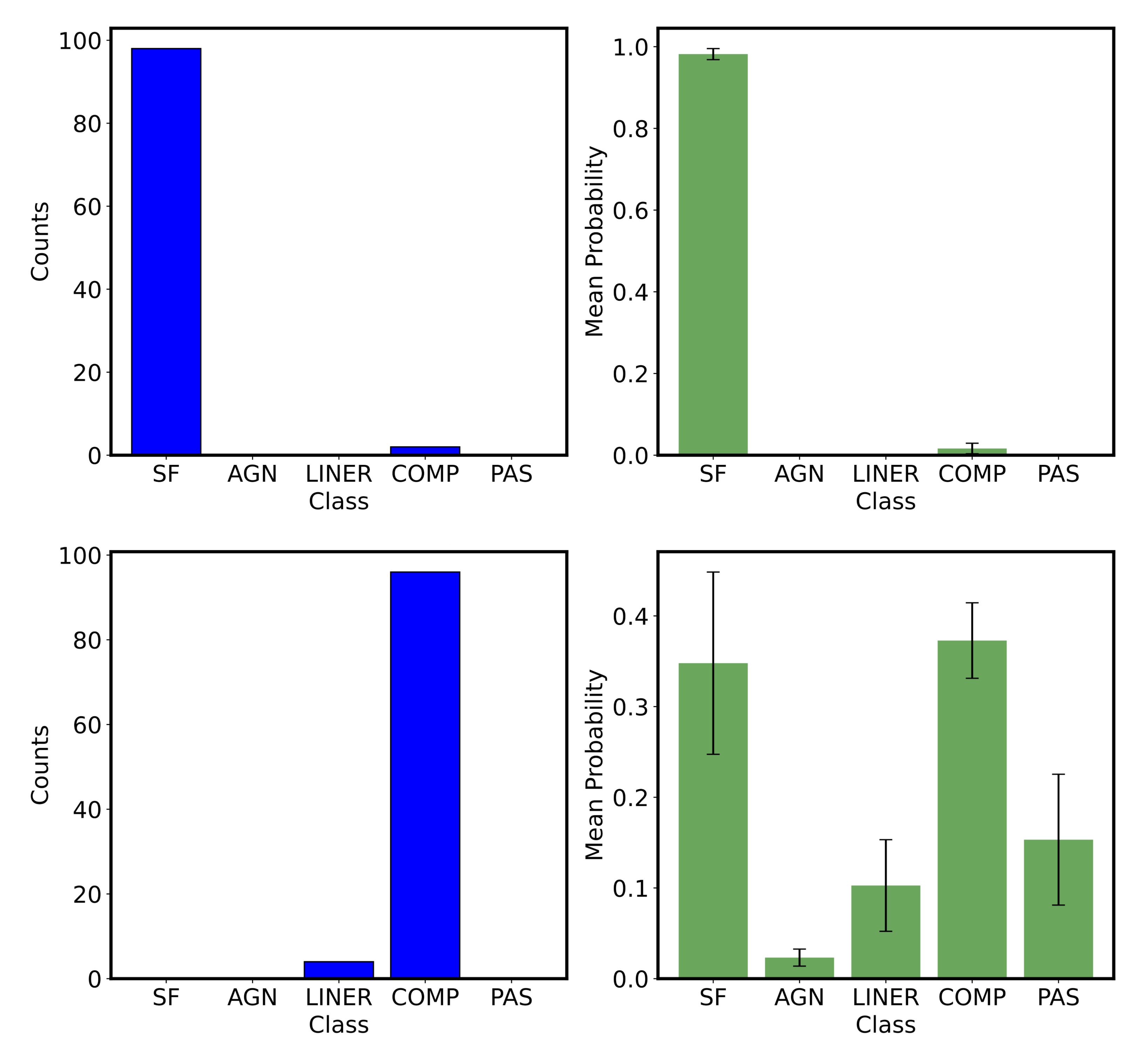}}
  \caption{Two examples of the classification output generated by our diagnostic, demonstrating a confident classification (top row) and a less confident classification (bottom row). Both objects have the same standard deviation (measurement error) in their EWs across all features. The output left (blue) histograms show: the resulting classifications based on Monte Carlo sampling the EW of the spectral features within their uncertainties, while the right (green) histograms show the corresponding probability for the different classes. The error bars represent the standard deviation of the predicted probabilities for each class. The classification result for the object depicted in the top row indicates a reliable classification, whereas the object in the bottom row exhibits ambiguous results.}
  \label{mc_classification_result}
\end{figure}

Figure \ref{mc_classification_result} shows two examples of the classification procedure accounting for measurement errors, as outlined in the preceding paragraphs. We consider two objects exhibiting similar standard deviations in their EWs across all features and we show the distribution of  (a) the predicted classes, and (b) the probability for each class. The top row demonstrates a classification with high confidence, where the majority of the classifications (left panel) are confined to a single class. The distributions of predicted probabilities (right panel) are dominated by this class, exhibiting low standard deviation. This object was identified as being near the center of its class distribution within the feature space. Conversely, the bottom row illustrates a classification with lower confidence, where the source is classified into multiple classes, although one dominates (left panel). The distribution of predicted probabilities (right panel) are dispersed across multiple classes, resulting in higher standard deviations (shown by the error-bars on the predicted probabilities) and ambiguous classification. This object lies near the intersection of multiple classes in the feature space, meaning that its classification is sensitive to small perturbations of its EW within the associated uncertainties. As a result, when the EW values are varied within their errors, the object may cross decision boundaries and transition into the locus of neighboring classes.

However, these probabilities should be treated with caution since they are very sensitive to the location of each object in the feature space with respect to the boundary between classes. This problem becomes particularly important for objects near the boundaries, and in the case of under-represented classes. Therefore, while the distribution of the predicted probabilities can provide insights into classification uncertainty, they cannot be used as a robust indicator. Instead we regard the distribution of the predicted classes as a more robust indicator of the classification confidence accounting for both measurement uncertainties and the intrinsic uncertainty of the classifier.

\subsection{Application on the HECATE catalog}

The Heraklion Extragalactic Catalog \citep[HECATE;][]{2021MNRAS.506.1896K,HECATEv2} is a comprehensive catalog that provides detailed information for 204,733 galaxies up to 200 Mpc. It is based on the HyperLEDA catalog and incorporates additional data from various extragalactic and photometric sources. The catalog offers comprehensive information, including positions, morphological information, multi-band photometry, distances, star-formation rates, stellar masses, gas-phase metallicities, and nuclear classifications. Activity classification for such a sample is important to estimate activity demographics, studies of the starburst and AGN connection, characterization of sources in wide-area multi-wavelength surveys, and transient events (e.g., tidal disruption events). Here, we utilized the revised HECATE catalog \citep[HECATEv2;][]{HECATEv2}. This version of HECATE includes spectroscopic classifications for 53,291 galaxies based on spectral line measurements available in the SDSS JHU-MPA DR8 catalog. These classifications are based on the diagnostic of \cite{2019MNRAS.485.1085S}. However, the number of galaxies with available spectra in SDSS has increased dramatically since DR8. As an application of this new method we obtain all available spectra from SDSS for the galaxies in the HECATE catalog from the DR17.

\begin{table*}[ht]
\centering
\caption{Comparison between the SoDDA diagnostic (ground truth) and our spectral classifier.}
\begin{tabular}{cccccccc}
\multicolumn{1}{l}{}                 & \multicolumn{7}{c}{SoDDA} \\
\multirow{7}{*}{\rotcell{This work}} &                & \multicolumn{1}{c}{SF} & \multicolumn{1}{c}{AGN} & \multicolumn{1}{c}{LINER} & \multicolumn{1}{c}{Composite}  & \multicolumn{1}{c}{Total}  \\ \hline
 & SF        & 16413 & 1393 & 403 & 2989 & 21198 \\
 & AGN       & 33 & 667 & 36 & 16 & 752 \\
 & LINER     & 10 & 25 & 447 & 84 & 566 \\
 & Composite & 190 & 23 & 179 & 802 & 1194 \\
 & Total     & 16646 & 2108 & 1065 & 3891 &  &   
\end{tabular}
\label{SoDDA_vs_SVM}
\end{table*}

The SDSS spectra were obtained from the DR17 using the \texttt{astroquery} Python package based on a cone search of 10 arcseconds around the galaxy coordinates. In total, we obtained spectra for 89,628 galaxies. We retained all spectra with reliable redshift measurements (i.e., \texttt{ZWARNING=0}), resulting in 88,514 spectra. Subsequently, our diagnostic is applied in all of these galaxies, and we perform two actions. First, we compare our classification results against the classification provided in the HECATE catalog, which were obtained using the SoDDA classifier (see Sect. \ref{gal_act_classes}). To achieve this, we adhere to all the quality criteria (S/N > 5 in all the aforementioned emission lines) that ensure reliable classification labels for the SoDDA diagnostic which results in 23,710 galaxies with highly reliable flux measurements in the H$\alpha$, H$\beta$, [\ion{O}{III}], [\ion{O}{I}], [\ion{N}{II}], and [\ion{S}{II}] lines. The results of this comparison are presented in Table \ref{SoDDA_vs_SVM}. These results are comparable to the performance metrics presented in Fig. \ref{conf_marix_principal}, suggesting that the classifier operates with high reliability. 

\begin{figure}[h]
  \centering
  \includegraphics[scale=0.35]{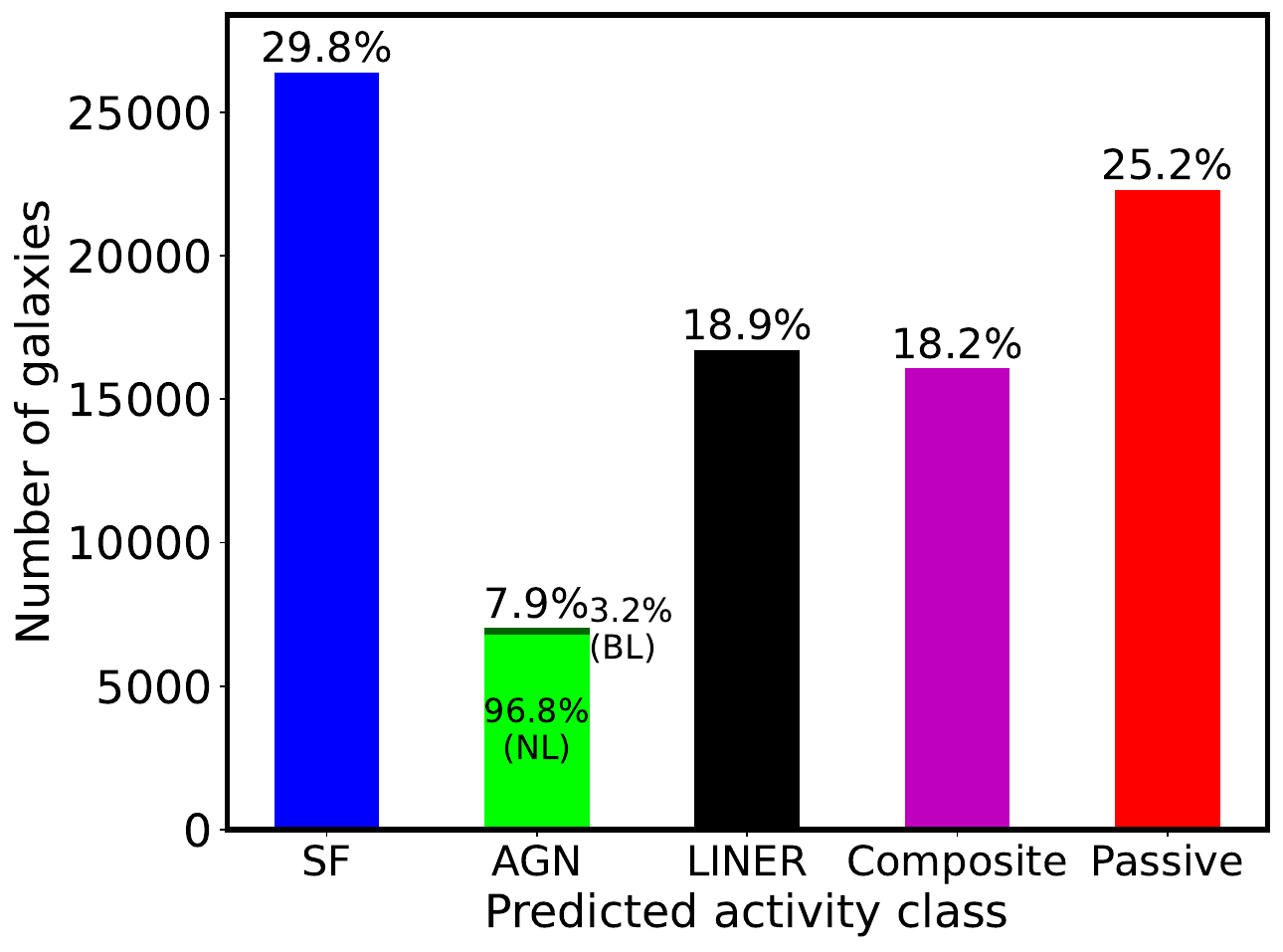}
  \caption{Distribution of activity classes derived from the implementation of our diagnostic on the HECATE catalog galaxies using SDSS DR17 spectra. Any galaxy classified as an AGN by our diagnostic is subsequently characterized as a broad-line (BL) or narrow-line (NL) AGN.}
  \label{pop_stat_HECATE}
\end{figure}

The second action is to implement our diagnostic to all galaxies in the HECATE catalog that have SDSS DR17 spectra (with \texttt{ZWARNING=0}), irrespective of whether they have classification labels from SoDDA. This will enable us to increase the population statistics per activity class of the HECATE catalog based on spectroscopic classifications, from 26\% to 43\%, include passive galaxies, and also identify broad- and narrow-line objects which currently are not characterized. We find that out of the 88,514 galaxies in HECATE, 26,393 (29,8\%) are star-forming, 7,024 (7.9\%) AGN, 16,721 (18.9\%) LINER, 16,075 (18.2\%) composite, and 22,301 (25.2\%) passive. AGNs are further subdivided into 6,800 narrow (96.8\%) and 224 broad-line (3.2\%) AGNs. These statistics are summarized in Fig. \ref{pop_stat_HECATE}. These classifications will be publicly available in a later release of the HECATE catalog.

We proceed by comparing the population statistics derived above (based on the spectra from DR17) for the galaxies in the HECATE catalog with those estimated from \cite{1997ApJ...487..568H}, which is regarded as a representative sample of galaxies in the local Universe. To improve the accuracy of the classifications of the \cite{1997ApJ...487..568H} sample, we apply on their emission-line measurements the SoDDA diagnostic \citep{2019MNRAS.485.1085S}, which incorporates three emission line ratios ([\ion{O}{III}]/H$\beta$, [\ion{N}{II}]/H$\alpha$, and [\ion{S}{II}]/H$\alpha$) simultaneously (the classification available in the original publication are based on the original version of the BPT diagnostics). Our analysis shows that the population statistics are broadly consistent, exhibiting only a 2-5\% deviation across all activity classes.

\subsection{Application to low-resolution spectra}

Our diagnostic was designed to be versatile and flexible so that it can be applied to spectra of different resolution than the ones used for its training (i.e., SDSS). We note that the resolution affects the classifier in two ways: (a) leakage of the emission lines into the continuum bands in poor resolution spectra affects the main activity classification since it will result in underestimation of the EW, and (b) affects the FWQM of the H$\alpha$+[\ion{N}{II}] blend which is used for the classification of AGN into narrow-line and broad-line objects. In order to minimize the leakage of the line flux into the continuum (even in the case of objects with broad lines; Sect. \ref{feat_selection}), the continuum is defined to be at least 40 \text{\AA} on either side from the center of the line band for the H$\alpha$+[\ion{N}{II}] and 15 \text{\AA} for H$\beta$ and the [\ion{O}{III}].

In order to test our diagnostic's performance on lower resolution and S/N spectra we first performed a simulation study for its application on spectra similar to those from the 6-degree field survey \citep[6dF;][]{2004MNRAS.355..747J,2009MNRAS.399..683J} and then we analyzed actual 6dF spectra. The 6dF galaxy survey, conducted using the 6dF spectrograph, is one of the most comprehensive surveys in the southern sky covering over 17,000 square degrees providing spectra for over 125,000 galaxies in the 4000–7500\,\text{\r{A}} range with a resolution of 5–8\,\text{\r{A}}.

We want to ascertain whether our EW measuring bands might result in flux leakage to the adjacent continuum in low resolution spectra, we applied our diagnostic on the objects from our test set of SDSS spectra (Sect. \ref{Data_sample}), after convolving them to match the resolution of the 6dF spectra. More specifically, given that the resolution of the 6dF and the SDSS spectra at an average wavelength is \text{FWHM}$_{\text{6dF}}$ = 6 $\AA$ and \text{FWHM}$_{\text{SDSS}}$ = 2.4 $\AA$, the convolution kernel is \text{FWHM}$_{\text{kernel}}$ = \text{FWHM}$_{\text{6dF}}^2$ - \text{FWHM}$_{\text{SDSS}}^2$ = 5.6 \text{\AA}. Subsequently, the EWs of all features in the test set were measured using the same bands (see Table \ref{EW_bands}). We find that the performance (recall score) using the smoothed spectra remained relatively unchanged across all classes (recall scores changed $\pm$1-2\% on average), with respect to the original SDSS spectra shown in Fig. \ref{conf_marix_principal}. This indicates that the line fluxes are well-contained within the defined measurement bands outlined in Sect. \ref{feat_selection} and that our diagnostic can be applied to lower resolution spectra without any modifications.

Next we apply our method to actual 6dF spectra which serve as a real-world example of spectra exhibiting significantly lower quality and resolution compared to those from the SDSS. In addition, the 6dF spectra are provided in uncalibrated photon counts and lack absolute flux calibration. Flux calibration of these spectra presents challenges due to calibration issues associated with instrumental and observational challenges that affect spectral quality \citep[e.g.,][]{2008MNRAS.386.1781P}. This in turn results in unreliable starlight subtraction, making the use of EW a good alternative.

In order to perform this exercise, we applied our diagnostic on all objects that are common to both SDSS and 6dF. Cross-matching our final sample (Sect. \ref{final_sample}) with the 6dF survey yields 1,093 common objects. Then, we measured the EWs for all discriminatory features from the 6dF rest-frame spectra. Table \ref{SDSS_vs_6df} presents the performance of our diagnostic when implemented on the 6dF spectra. For this comparison, we consider as ground truth the classification from SoDDA (Sect. \ref{gal_act_classes}). We observed a reduction in performance, which is expected due to generally poorer quality of the 6dF spectra. Furthermore, we find that when we exclude objects with comparable predicted probabilities of belonging to more than one class, indicating unreliable classifications (Sect. \ref{classification_confidence}) the fraction of correct classifications reaches 60\%. However, it is important to note that the number of available objects used for this test is small for reliable statistics. Although there is a reduction in performance, due to the lower quality of the 6dF spectra, the diagnostic remains a robust method for classifying galaxy spectra across various surveys considering the intrinsic limitations of each survey. 

Based on the results of the simulation study, the reduced performance cannot be only attributed to differences in the SDSS and 6dF spectral resolution. In addition, the fibers used to obtain the 6dF spectra have a size of 6.7 arcsec (in comparison to $\sim$3 arcsec for SDSS), also leading to aperture effects. Aperture bias poses a challenge because it dictates the balance of nuclear versus host-galaxy light in a spectrum. While we might expect larger apertures to simply dilute AGN signatures, \cite{2014MNRAS.441.2296M} showed that the reality is far more complex, often driven by the inclusion of circumnuclear star-forming regions, which have a wide range of relative emission line strengths depending on the age and the metallicity of their stellar populations. This unpredictability means that we cannot simply assume a systematic trend across the sample, requiring the assessment of aperture effects on a case-by-case basis.

\begin{table*}[ht]
\centering
\caption{Comparison between classifications based on measurements performed on the 6dF and corresponding SDSS spectra.}
\begin{tabular}{ccccccccc}
\multicolumn{1}{l}{}                 & \multicolumn{7}{c}{SDSS} \\
\multirow{7}{*}{\rotcell{6dF}} &                & \multicolumn{1}{c}{SF} & \multicolumn{1}{c}{AGN} & \multicolumn{1}{c}{LINER} & \multicolumn{1}{c}{Composite}  & \multicolumn{1}{c}{Passive} & \multicolumn{1}{c}{Total} \\ \hline
 & SF        & 163 & 7  & 15 & 98 & 13  & 296 \\
 & AGN       & 4   & 38 & 14 & 5  & 5   & 66 \\
 & LINER     & 4   & 1  & 33 & 13 & 14  & 65 \\
 & Composite & 35  & 4  & 16 & 81 & 2   & 138 \\
 & Passive   & 0   & 0  & 32 & 8  & 401 & 441 \\
 & Total     & 206 & 50 & 110 & 205 & 435  &   
\end{tabular}
\label{SDSS_vs_6df}
\end{table*}

Spectra obtained from the 6dF survey have previously been employed to classify galaxy activity using the BPT diagnostics \citep{2019ApJ...872..134Z}. In that study, two catalogs of galaxies are provided: one containing optically identified AGN (based on BPT diagnostics; 3,365 objects) and the other containing galaxies that do not exhibit AGN activity (6,787 objects) which are separated into objects that show emission lines (3,837 objects) and non-emission line objects (2,950 objects). 

As an application of our method we apply it on the set of objects with classifications from \cite{2019ApJ...872..134Z}. We retrieve the spectra for these two galaxy catalogs from the {6dF}\footnote[2]{\url{http://www-wfau.roe.ac.uk/6dFGS/}} using their celestial coordinates. Then we classify these spectra using our diagnostic. Based on our diagnostic from the AGN sample reported in \cite{2019ApJ...872..134Z} we identify 265 (7.9\%) galaxies as AGN, 598 (17.8\%) as star-forming, 770 (22.9\%) as LINER, 1,052 (31.3\%) as composite, and 680 (20.2\%) as passive. The results for the objects showing emission lines from the non-AGN catalog are as follows: 635 (16.5\%) star-forming, 30 (0.8\%) AGN, 606 (15.8\%) LINER, 713 (18.6\%) composite, and 1,853 (48.3\%) passive. The results for the objects with no presence of emission lines from the non-AGN catalog are as follows: 43 (1.5\%) star-forming, 8 (0.3\%) AGN, 259 (8.8\%) LINER, 144 (4.9\%) composite, and 2,496 (84.6\%) passive. We see that for the AGN sample of galaxies from \cite{2019ApJ...872..134Z} we identify 7.9\% of the galaxies as actual AGNs. Also, the LINER and composite objects, this fraction reaches 54\%. 
However, it should be noted that measuring emission line ratios from the 6dF spectra is prone to several biases due to poor spectral resolution and the lack of flux calibration which affects the starlight subtraction and of course the measurement of the emission lines themselves. In addition to these effects, these discrepancies may stem from the low quality of the 6dF spectra and the flux calibration that suffers from various artifacts, as previously discussed and which affects the measurement of absolute line fluxes. These factors hinder the accurate measurement of emission line fluxes making the classifications from \cite{2019ApJ...872..134Z} unreliable.

\subsection{The effect of the stellar continuum}

One potential limitation of our method is that it does not require removal of the stellar component of galaxy spectra (starlight subtraction). In general, starlight subtraction is used in order to measure the intrinsic flux of emission lines, especially those that are buried in strong stellar absorption features, which may lead to underestimation of their intensity. However, starlight subtraction also presents significant challenges. One major drawback is the over or under-subtraction of the starlight resulting in potential contamination or residuals, which can lead to inaccuracies in derived quantities if not carefully handled \citep[see][for a discussion on the sensitivity of line measurements on the starlight subtraction process]{2014MNRAS.441.2296M}. Furthermore, assumptions about the stellar populations, such as the initial mass function or the adoption of particular stellar libraries can introduce uncertainties \citep{2009ApJ...699..486C}. In addition, in the case of AGN, starlight subtraction may give wrong estimates for the emission line fluxes since it models the AGN continuum as stellar continuum affecting the stellar absorption lines and hence the estimated intrinsic flux of the emission lines of interest. This is particularly important for identifying low luminosity AGN especially in early type galaxies where the absorption lines may be masked by the stellar component \citep{2002ApJ...579..545C,2017A&A...604A..99C}. 

\begin{figure}[h]
  \centering
  \includegraphics[scale=0.4]{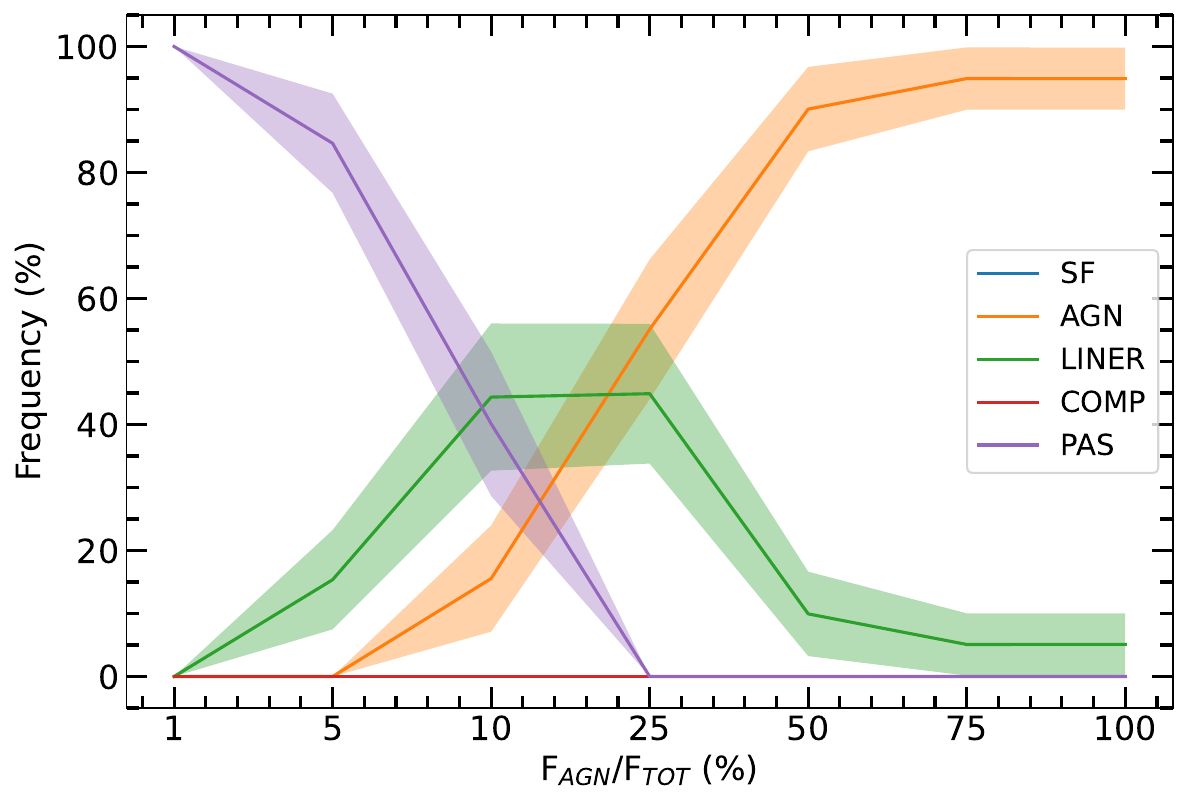}
  \caption{Classification frequencies of a composite spectrum, derived from combining AGN and passive spectra, with respect to the classes considered in our diagnostic scheme, as a function of AGN fractional contribution to the total nuclear (3$^{\prime\prime}$) continuum in the SDSS $g$-band. We observe a transition from passive to LINER and then AGN galaxy classification when the AGN contribution to the nuclear flux in the SDSS $g$-band reaches approximately 10\% and 25\% respectively indicating that our diagnostic is sensitive to AGN contributions in passive galaxies as low as 10\% in the $g$-band. The curves depict the average predictions of each class for 20 randomly generated composite spectra while the shaded areas represent the standard deviation of the classification frequencies.}
  \label{AGN_PAS}
\end{figure}

Our diagnostic tool avoids these biases, as it eliminates the need for starlight subtraction from the spectra. However, this presents another challenge: the suppression of characteristic spectral features that may be concealed within a dominant stellar continuum, resulting in low luminosity AGN being misclassified into other classes (e.g., composite) or even passive.

To quantify the impact of the stellar continuum on the classification of star-forming and AGN galaxies, we apply the diagnostic on simulated spectra of composite galaxies with known AGN contribution. First we select galaxies with the same apparent magnitude in $g$-band from each of our final samples (Sect. \ref{final_sample}) of: star-forming, AGN, and passive galaxies. The $g$-band used here is derived from the SDSS 3$^{\prime\prime}$ fiber (fiberMag) which corresponds to the nucleus of a galaxy. In addition, this ensures consistency with the spectroscopic aperture used for the ground truth classification (SoDDA) minimizing aperture-related effects. The AGNs utilized for this analysis have a median luminosity of the H$\alpha$ line $L_{\mathrm{H}\alpha} \approx 10^{41} \ \mathrm{erg ~s^{-1}}$, which is generally regarded as a moderate luminosity for an AGN making it ideal for low luminosity AGN. Then, composite galaxy spectra are created by combining randomly drawn spectra of AGN, star-forming and passive galaxies. This way we consider both types of composite spectra: AGN + passive and AGN + star-forming. In total, in this exercise, we use spectra from 460 star-forming, 42 AGNs, and 153 passive galaxies with 18.50 < $m_g$ < 18.55 (AB mag), ensuring spectra of similar quality. In order to simulate different AGN contributions we keep the AGN spectrum as is and we multiply the other (star-forming or passive, depending on the analysis) spectrum by a constant $c$. Then the AGN contribution (in $g$-band continuum) to the composite (total) spectrum is calculated as $1/(1+c)$ and we select appropriate values for $c$ to simulate AGN fractional contributions of 1, 5, 10, 25, 50, 75, and 100\%. In total, 20 composite spectra for each AGN contribution fraction are produced, by randomly drawing spectra from each of the considered types (AGN, passive, star-forming). The resulting composite spectra are then classified with our diagnostic tool and we calculate the average and standard deviation of the predictions of each composite spectrum to belong to each one of the considered classes in our classification scheme. 

Figure \ref{AGN_PAS} presents the average predictions (classification frequency) per activity class for the composite spectra of AGN + passive as a function of the AGN fractional contribution in the nuclear (3$^{\prime\prime}$) $g$ band continuum. This is relevant for AGN in elliptical galaxies or galaxy bulges. This plot illustrates that even without removing the stellar component our diagnostic is able to identify passive galaxies hosting AGN with contribution as low as 25\% in the $g$ band. For lower AGN contribution (below 25\% to 10\%), our composite spectra are classified as LINERs with AGN being the second ranking class.

\begin{figure}[h]
  \centering
  \includegraphics[scale=0.4]{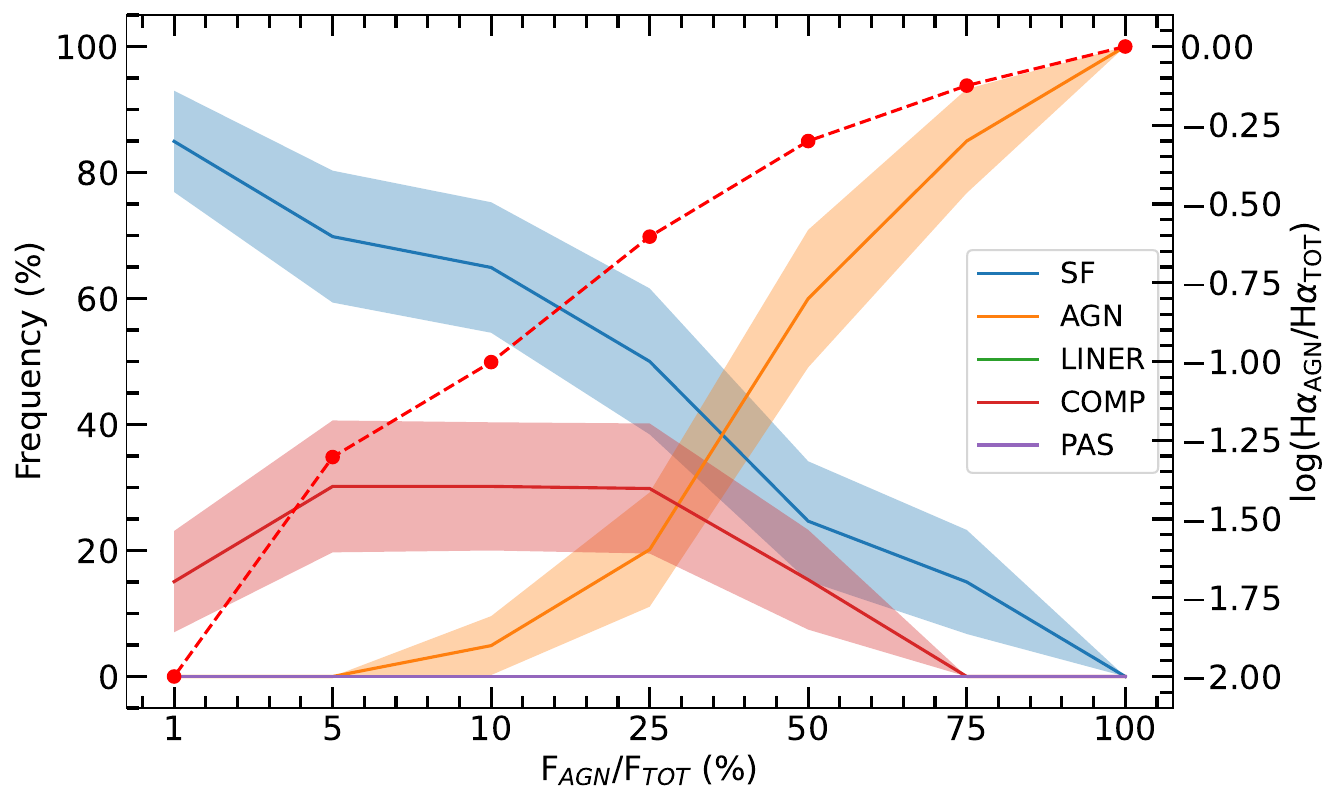}
  \caption{Similar analysis, as in Fig. \ref{AGN_PAS}, is but for composite spectra combining star-forming and AGN spectra. The red dashed line depicts the logarithm of the ratio of the H$\alpha$ luminosity of the AGN to the total H$\alpha$ luminosity of the composite spectra. Although the transition from star-forming to AGN occurs when the AGN fractional contribution to the nuclear SDSS $g$-band continuum approaches $\sim$40\%, the corresponding H$\alpha$ fractional contribution is $\sim$30.}
  \label{SFG_AGN}
\end{figure}

In Fig. \ref{SFG_AGN} we present the same analysis, but for composite spectra of AGN + star-forming galaxies. The primary objective of this exercise is to assess the diagnostic’s sensitivity in scenarios where star formation is present alongside weak AGN emission. This could result in misclassifications of weak AGNs as star-forming galaxies, which will contaminate our classifications for star-forming galaxies and lead to the omission of AGNs. We see that the identification of an AGN in the composite spectra of AGN + star-forming is possible when the AGN contribution in the nuclear $g$-band continuum becomes more that $\sim$40\%. However, when we consider the contribution of the AGN in the H$\alpha$ luminosity to the total composite spectrum (red line and right y-axis of Fig. \ref{SFG_AGN}) we can detect AGN at the 30\% contribution level. This is a good score considering the fact that the AGN spectra used in the construction of the composite spectra also include their host galaxy star-forming component which has a non negligible contribution given the moderate AGN H$\alpha$ luminosities of $\sim$1$0^{41}$ $\mathrm{erg ~s^{-1}}$. This star-forming component alongside the “host galaxy” component will both contribute in the dilution of the AGN emission, especially in the case of the H$\alpha$ + [\ion{N}{II}] lines. From the same figure we also see that composite spectra classified as star-forming with a classification frequency exceeding $\sim$60\% are indeed dominated by star-formation with small AGN contribution (i.e., the AGN fraction in the $g$-band continuum is below $\sim$25\%, and even lower in the H$\alpha$ flux). Similarly, galaxies with classification frequency higher than 50\% are AGN dominated. Additionally, we observe that as the AGN fraction increases in the composite spectra of AGN + star-forming galaxies, the second most likely class is composite galaxy. At the transition point, where the AGN fraction is approximately 30\%, this classification frequency competes with both AGN and star-forming classifications. This outcome is expected, as composite galaxies can result from moderate-luminosity AGNs within galaxies that exhibit substantial star formation.

\begin{figure}[!h]
  \centering
  \includegraphics[scale=0.35]{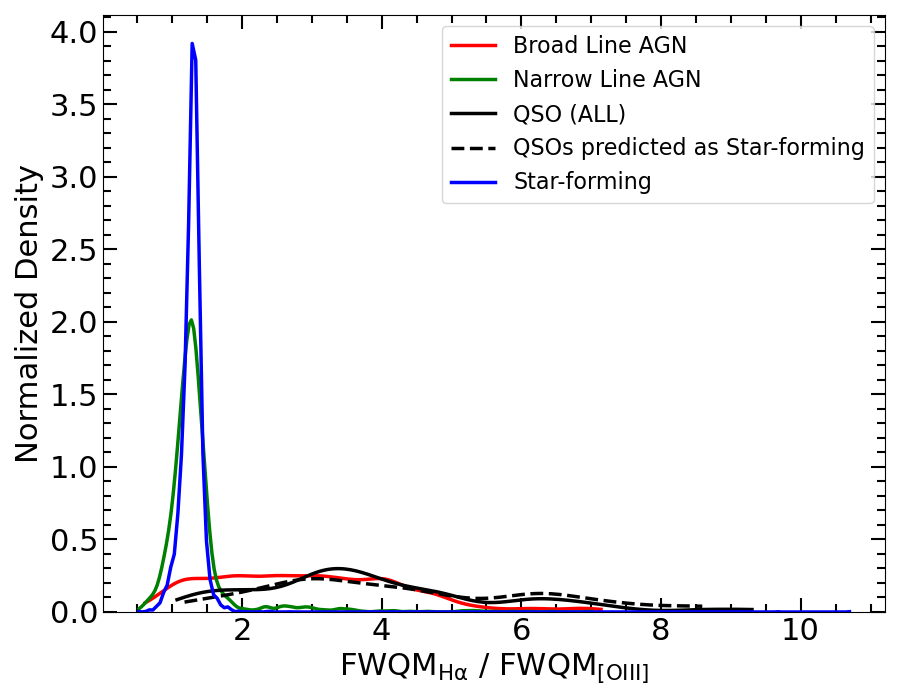}
  \caption{Histogram of the ratio of the FWQM of the H$\alpha$ to the [\ion{O}{III}] $\lambda$5007\AA\, lines for the star-forming, broad-line and narrow-line AGN in the training sample (blue, red, and green lines respectively). We see that the distributions for the narrow-line AGN and star-forming galaxies are almost identical, while the broad-line AGN extend to much larger FWQM ratio, with only a small fraction in the region of the star-forming galaxies or narrow-line AGN. We also show with the black lines the very broad-line objects (QSO-like spectra) classified as AGN (solid line) and star-forming (dashed line). We see that the majority of the QSO-like objects classified as star-forming have much larger FWQM ratio than bona-fide star-forming galaxies, allowing us to use the FWQM ratio as a metric to correctly classify them.}
  \label{FWQM_ratio}
\end{figure}

\subsection{Sensitivity to flux calibration} \label{flux_cal_sens}

Accurate flux calibration is a fundamental prerequisite for reliably measuring emission-line fluxes and performing starlight subtraction. In certain instances, flux calibration can be challenging due to the instrument or survey design. Our diagnostic overcomes this difficulty since it uses EWs which make it insensitive to the absolute flux  or even spectrograph sensitivity calibration. To evaluate the sensitivity of our diagnostic to flux calibration effects and its applicability in such situations, we conducted the following test. We multiplied the SDSS spectra from the test set (Sect. \ref{final_sample}) with (a) polynomial functions of up to fifth degree, or (b) sine function. This approach was designed to simulate systematics that could occur in cases of problematic flux/sensitivity calibration. Then, we applied our diagnostic (Sect. \ref{Developing_the_diagnostic_tool}) on these transformed spectra. The performance scores across all classes remained unchanged (recall scores changed $\pm$1-2\% on average), as a result of the relatively narrow spectral windows used for measuring the EWs of the spectral lines (see Sect. \ref{feat_selection}) and the fact that we are using EW instead of line fluxes. Because of the narrow range of the continuum and line bands the linear function for the calculation of the continuum under a spectral line is a good approximation. For more details about the specifics of this exercise, please refer to Appendix \ref{AppB}.

\subsection{Distinguishing broad-line AGN from other sources of line broadening}

In Sect. \ref{FWQM_BLR}, we introduced an additional classification step to indicatively identify broad-line AGN using only the FWQM of H$\alpha$ + [{\ion{N}{ii}}]. Although this approach is highly effective (see Sect. \ref{perf_on_BLR}), it may include broad-line objects resulting from shocks or outflows rather than from a classical broad-line region. The two types of broadening can be discriminated by comparing the width of the Balmer and the forbidden lines, since in a genuine broad-line AGN only the Balmer lines are broad  \citep[see e.g.][]{2025A&A...696A.133K}.

To test this we consider the ratio of the FWQM of the H$\alpha$ to the [\ion{O}{III}] line. We use the H$\alpha$ line for its higher S/N and weaker of stellar absorption relative to H$\beta$. While the [\ion{N}{II}] lines may result in artificially larger FWQM, we find that this is not the case since there is a strong linear correlation between the H$\alpha$/[\ion{O}{III}] ratios of FWQM and their line width ($\sigma$) ratios. Figure \ref{FWQM_ratio} compares the H$\alpha$/[\ion{O}{III}] FWQM ratio distributions for broad-line and narrow-line AGNs. We see that the vast majority of narrow-line AGNs have FWQM ratios less that $\sim$1.5 (with a median value of $\sim$1.25), while the vast majority of broad-line AGN have higher ratios. We find that the broad-line AGN with FWQM ratios of $\sim$1.25, have median FWHM for both the H$\alpha$ and the [\ion{O}{iii}] line of $\sim$850 km s$^{-1}$, consistent with line broadening driven by shocks or outflows rather than a classical broad line region. However, broad-line AGNs with FWQM ratios $>$1.25 are consistent with the definition of a broad-line AGN (median FWHM of H$\alpha$ $\gtrsim$ 1000 km s$^{-1}$ and narrower [\ion{O}{iii}] with $\lesssim$ 300 km s$^{-1}$). Thus, a criterion of the FWQM ratio of H$\alpha$ to [\ion{O}{III}] of $<1.75$ is effective in separating bona-fide broad-line AGNs against shocks/outflows.

To further investigate how our diagnostic handles genuine cases of broad-line AGN with lines much broader than the $\sim$470 km s$^{-1}$ limit used for the definition of our sample (Section \ref{def_AGN_subclasses}) we used quasi-stellar objects (QSOs) to perform the following exercise. We selected a local sample ($z<$ 0.1) of good quality (\texttt{ZWARNING=0} and \texttt{snMedian}\footnote[3]{\url{https://skyserver.sdss.org/dr15/en/help/browser/browser.aspx\#\&\&history=description+SpecObjAll+U}} $>$ 10) QSOs from SDSS (\texttt{"SUBCLASS"="QSO"}) to apply our diagnostic. We find that 65\% of the QSOs are classified as AGN, 27\% as star-forming, and 8\% as composite. The missclassification to star-forming galaxies is because our bands are too narrow to fully encompass these very broad Balmer lines. This effect can be remedied by imposing a threshold based on the FWQM ratio (> 1.75) of the H$\alpha$ to [\ion{O}{III}] lines as above, since the broad-line objects will have much higher ratios than the star-forming galaxies 
(Fig. \ref{FWQM_ratio}).

\section{Conclusions} \label{Conclusions}

In this work, we introduced a new activity diagnostic tool that includes all possible types of galaxy activity under one unified classification scheme utilizing only the EW of the H$\beta$, [\ion{O}{III}]\,$\lambda$5007, and the H$\alpha$ + [\ion{N}{II}]\,$\lambda$$\lambda$6548,84 spectral lines. Our diagnostic method offers significant improvements over similar studies that employ EW as a classification tool \citep[e.g.,][]{2010MNRAS.403.1036C}. It entirely eliminates the need for measuring emission line fluxes and introduces a more comprehensive activity classification framework, which notably includes the often-overlooked category of passive galaxies. Our results are summarized as follows.

\begin{enumerate}
  \item Our new diagnostic tool introduces a classification scheme that categorizes galaxies into the main five activity classes: star-forming, AGNs, LINERs, composite, and passive galaxies under one unified scheme. Notably, our diagnostic tool achieves near-perfect scores across all classes.
  \item AGN galaxies with high resolution spectra can be further classified into two subclasses: broad and narrow-line AGNs, both of which exhibit high completeness with only one feature (i.e., the FWQM of the H$\alpha$ + [\ion{N}{II}] blend).
  \item A significant advantage of our diagnostic is its direct applicability to uncalibrated optical spectra, thereby eliminating the need for flux calibration, removal of the stellar component, and fitting or deblending of the spectral line profiles. The latter is particularly important in the case of the H\(\alpha\) and [\ion{N}{ii}] doublet lines, an often challenging and model-dependent process. Moreover, it is largely insensitive to reddening effects, ensuring reliable classification even for spectra affected by significant dust attenuation.
  \item Genuine broad-line AGN and broad-line objects due to outflows or shocks can be discriminated based on the comparison of the FWQM of the H$\alpha$ and the [\ion{O}{iii}] $\lambda$5007 lines, a feature that is included in our pipeline.
  \item AGN hosted in passive galaxies with an AGN contribution greater than $\sim$25\% in the $g$-band continuum are recovered even without starlight subtraction. Similarly, AGN situated within star-forming galaxies can be reliably identified when their nuclear $g$-band contribution is as low as $\sim$40\%. 
  \item The narrow spectral range window (4864-6584\,\text{\r{A}}) of our diagnostic, its high completeness, and its independence from the resolution of the spectrograph makes our diagnostic extremely versatile and enablesthe classification of high-redshift galaxies such as those observed with the James Webb Space Telescope (JWST) and the Dark Energy Spectroscopic Instrument \citep[DESI;][]{2016arXiv161100036D}, making it an invaluable tool for studying galaxy activity across the universe.
 \end{enumerate}

\section{Data availability} \label{sec7}

The code, including detailed documentation and usage instructions, is publicly available via a GitHub repository\footnote[4]{\url{https://github.com/BabisDaoutis/OptSpecClassifier}}.

\begin{acknowledgements} 

We thank the anonymous referee for their insightful suggestions that helped to clarify and strengthen this work. CD and EK acknowledge support from the Public Investments Program through a Matching Funds grant to the IA-FORTH. The research leading to these results has received funding from the European Research Council under the European Union’s Seventh Framework Programme (FP/2007-2013)/ERC Grant Agreement n. 617001, the European Union’s Horizon 2020 research and innovation programme under the Marie Skłodowska-Curie RISE action, Grant Agreement n.873089 (ASTROSTAT-II), and the Smithsonian Astrophysical Observatory Predoctoral Program with funding under the NASA grant 80NSSC21K0078. KK acknowledges support by the institutional project RVO:67985815 and by the INTER-COST LUC24023 project of the INTER-EXCELLENCE II programme of the Czech Ministry of Education, Youth and Sports. Funding for the Sloan Digital Sky Survey IV has been provided by the Alfred P. Sloan Foundation, the U.S. Department of Energy Oﬃce of Science, and
the Participating Institutions. SDSS-IV acknowledges support and resources
from the Center for High Performance Computing at the University of Utah.
The SDSS website is www.sdss.org. SDSS-IV is managed by the Astrophysical Research Consortium for the Participating Institutions of the SDSS Collaboration including the Brazilian Participation Group, the Carnegie Institution for Science, Carnegie Mellon University, Center for Astrophysics | Harvard \& Smithsonian, the Chilean Participation Group, the French Participation Group, Instituto de Astrofísica de Canarias, The Johns Hopkins University, Kavli Institute for the Physics and Mathematics of the Universe (IPMU)/University of Tokyo, the Korean Participation Group, Lawrence Berkeley National Laboratory, Leibniz Institut für Astrophysik Potsdam (AIP), Max-Planck-Institut für Astronomie (MPIA Heidelberg), Max-Planck-Institut für Astrophysik (MPA Garching), Max-Planck-Institut für Extraterrestrische Physik (MPE), National Astronomical Observatories of China, New Mexico State University, New York University, University of Notre Dame, Observatário Nacional/MCTI, The Ohio State University, Pennsylvania State University, Shanghai Astronomical Observatory, United Kingdom Participation Group, Universidad Nacional Autónoma de México, University of Arizona, University of Colorado Boulder, University of Oxford, University of Portsmouth, University of Utah, University of Virginia, University of Washington, University of Wisconsin, Vanderbilt University, and Yale University.

\end{acknowledgements}

%
%






   
  



\bibliographystyle{aa}
\bibliography{references}

\begin{appendix} 

\section{Optimization of the SVM hyperparameters} \label{AppA}

An SVM model is specified by at least three significant hyperparameters: the ''tolerance'' C (regulating the allowed mis-classifications), the kernel type (for the calculation of the dot product in the transformed feature space), and the kernel hyperparameters themselves (i.e., the constants in the kernel’s functional form). In any implementation, these come with default values, but in practice they need to be optimized for the data at hand.

In the pursuit of optimizing parameters for machine learning models such as SVM, randomized search emerges as a robust alternative to conventional to static methods like grid search. Randomized search is a methodology that explores a predetermined number of random combinations of the parameters. This approach facilitates the efficient identification of optimal parameters by better exploring the parameter space and surpassing exhaustive search techniques in terms of computational efficiency.

This process begins by defining parameter distributions for each hyperparameter, enabling random selection and evaluation of candidates from the parameter space. Unlike grid search, which assesses all possible combinations, randomized search evaluates a random subset. After identifying potential value ranges for each hyperparameter, multiple random parameter sets are generated, each used to train the model. Performance metrics such as accuracy, recall, or F1-score are computed, and the set with the best performance is chosen as the optimal configuration.

\begin{table}[h]
\caption{Hyperparameter distribution for the primary diagnostic tool based on the SVM algorithm.} 
\centering 
\begin{tabular}{l c c}
\hline\hline
Parameter & Search values & Best value \\ 
\hline
\texttt{C} & Uniform(1, 1000) & 100 \\ 
\texttt{kernel} & \texttt{linear}, \texttt{rbf}, \texttt{poly}, \texttt{sigmoid} & rbf \\ 
\texttt{gamma} & Uniform(0.01, 100), \texttt{scale}, \texttt{auto} & scale \\ 
\hline
\end{tabular}
\tablefoot{For a comprehensive explnation of all parameters see scikit-learn's {documentation.}\footnote[1]{\url{https://scikit-learn.org/stable/modules/svm.html}}}
\label{table:svm_hyperparams} 
\end{table}

\section{Insensitivity to flux calibration} \label{AppB}

Flux calibration is a critical step in the processing of optical spectra, directly impacting the accuracy of emission line flux measurements in galaxies. It involves correcting the raw observed spectra for the instrumental response and atmospheric transmission, thereby translating counts or relative fluxes. Without proper flux calibration, systematic biases may be introduced in the derived emission line fluxes, particularly when comparing lines across different wavelength regions (e.g., H$\beta$ and H$\alpha$), which can lead to significant errors in key diagnostics such as dust attenuation, gas-phase metallicity, and star formation rates. Even moderate calibration errors can propagate non-linearly into derived quantities, especially in line ratio diagnostics that rely on lines widely spaced in wavelength. Therefore, careful attention to flux calibration procedures, is indispensable for robust emission line analyses and the physical interpretation of galaxy spectra.

EWs offer a robust alternative to absolute flux measurements in cases where flux calibration is uncertain or unreliable. By construction, the EW of an emission line is the ratio of the line flux to the adjacent continuum flux density, making it a quantity that is inherently insensitive to multiplicative calibration errors that affect both the line and continuum equally. When the wavelength range is short and the continuum can be approximated with a linear function the EWs remain relatively stable even when the absolute spectral shape or throughput is poorly characterized.

\begin{figure}[h]
  \resizebox{\hsize}{!}{\includegraphics{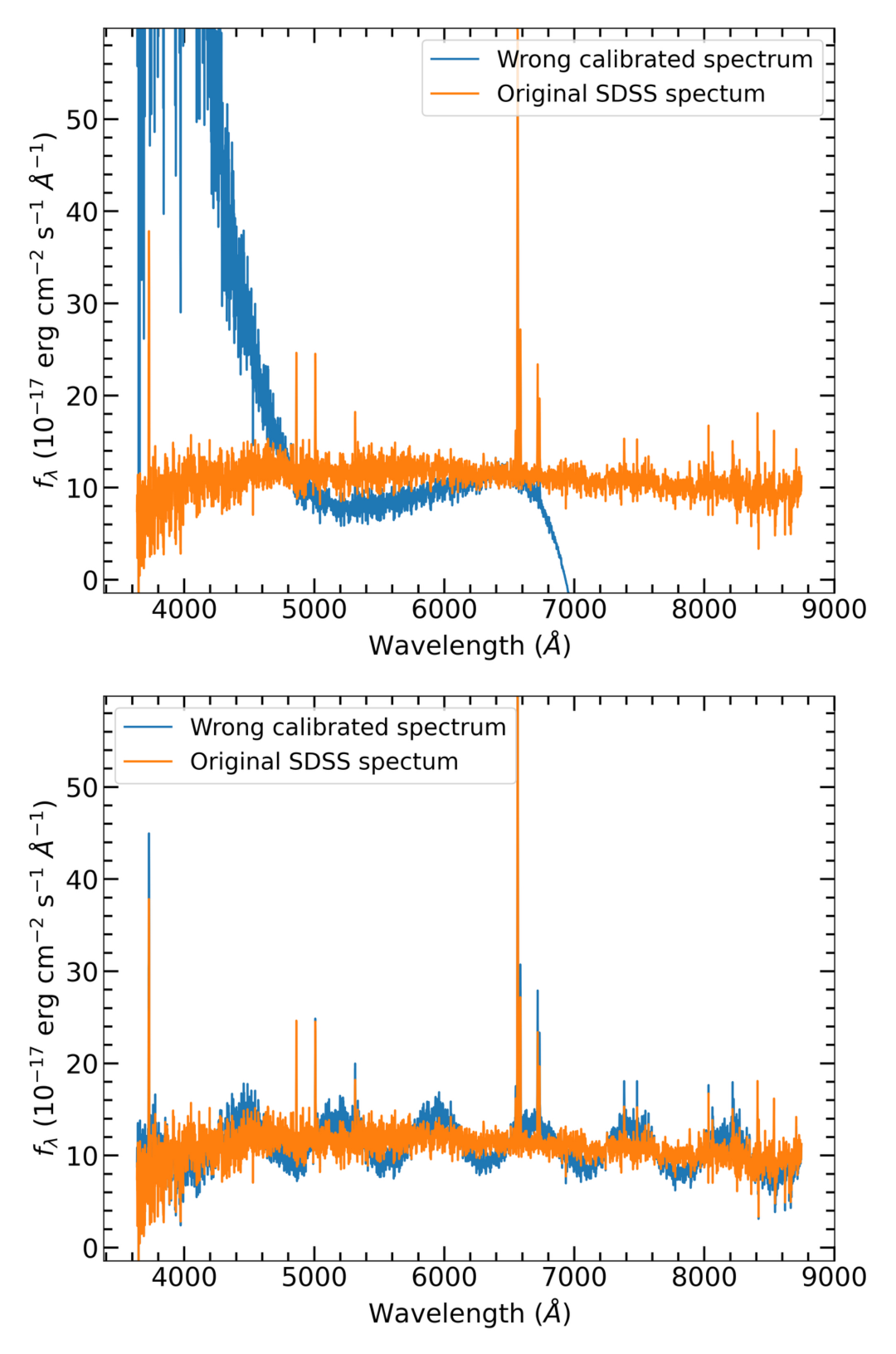}}
  \caption{Examples of spectra used to assess the impact of erroneous flux calibration. The orange spectrum indicates the original spectrum, while the blue spectra show the same spectrum convolved with a fifth order polynomial (top) and a sine function (bottom). Notably, both convolved spectra were accurately classified by our diagnostic, yielding the same classification as when utilizing the original SDSS spectra.}
  \label{wr_fl}
\end{figure}

Consequently, we assess the sensitivity of our diagnostic in instances where the flux calibration may have been executed inadequately. To accomplish this, we employed the full set (see Sect. \ref{Data_sample}) of spectra from the test and convolved them with a fifth-degree polynomial. Subsequently, we repeated the EW measurements of our discriminating features on these transformed spectra. Then, our diagnostic was applied to the EWs measured from the transformed spectra. We found that all the performance metrics (e.g., recall and accuracy) remained unchanged across all classes (recall scores changed $\pm$1-2\% on average). The above analysis was repeated but the test set spectra were convolved with a sine function to simulate a scenario where the flux calibration generates periodic patterns on the spectra. Once again, we obtained similar results. Figure \ref{wr_fl} shows an example of a SDSS spectrum after it has been convolved with the fifth-degree polynomial (top) and the sine function (bottom).

\section{Measuring EWs} \label{AppC}

\begin{figure}[h]
  \resizebox{\hsize}{!}{\includegraphics{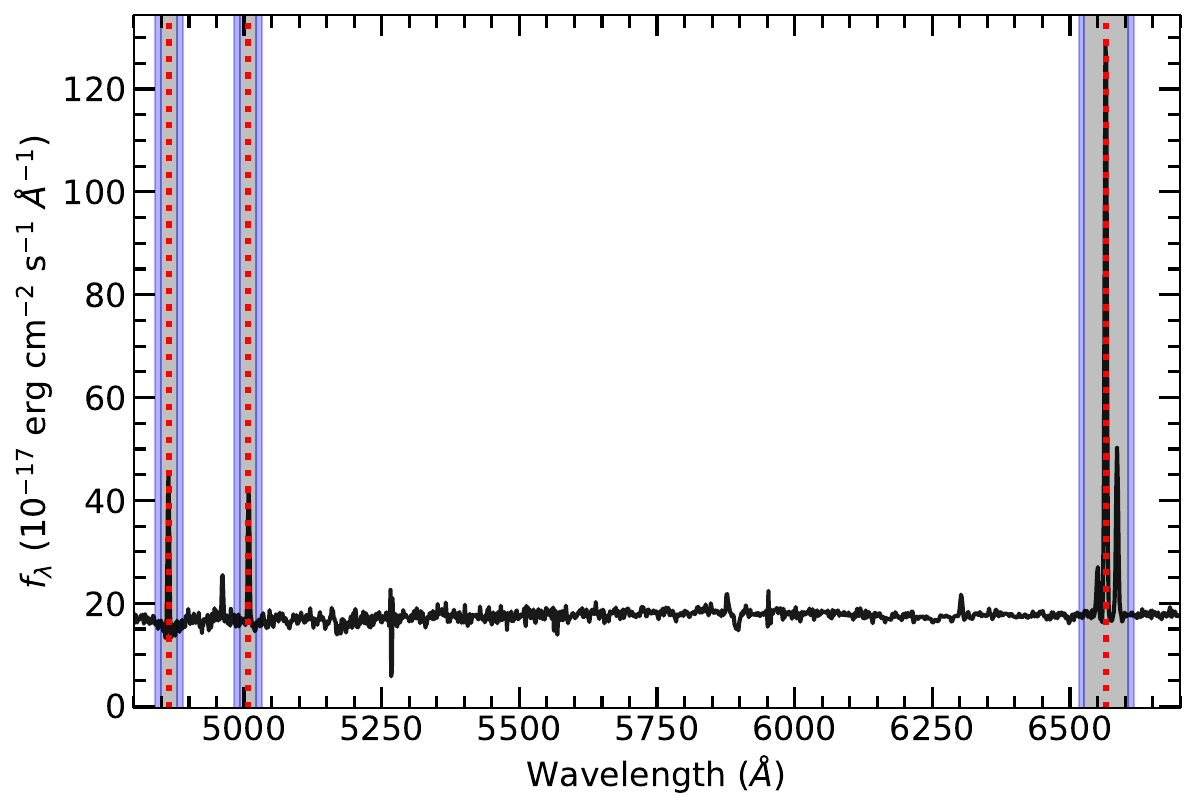}}
  \caption{Flux density versus wavelength for an example SDSS spectrum (black) to show the location of our discriminating features along with their respective wavelength ranges (Table \ref{EW_bands}). The central wavelengths of the targeted features are indicated by red dashed lines, and the shaded gray areas represent the ranges for the EW calculation of each spectral feature. The purple shaded areas n the either side of each band show the areas used to estimate the continuum for the liner fit.}
  \label{spectrum_line_ranges}
\end{figure}

To measure the EWs of the targeted features (see Table \ref{EW_bands}), each spectrum was first corrected for redshift using spectroscopic redshifts reported in the SDSS. We then defined fixed spectral windows centered on the target lines-such as H$\beta$, [\ion{O}{III}], and the H$\alpha$ + [\ion{N}{II}] $\lambda \lambda$6548,84 blend avoiding contamination from neighboring lines. For each spectral feature, the local continuum was estimated by fitting a line to adjacent line-free regions starting from the edges of the spectral windows used for the spectral features (see Table \ref{EW_bands}) by fitting a straight line on the flux density as a function of wavelength using the data from the blue and red continuum bands for each line. The equation for estimating the continuum is the following: 

\begin{equation}
F_{\lambda,\mathrm{cont}} = \alpha \lambda + \beta,
\label{cont_eq}
\end{equation}
where the $\alpha$ is the slope and is $\beta$ the intercept. All EWs was calculated as:

\begin{equation}
    \mathrm{EW} = \int_{\lambda_c -\Delta x/2}^{\lambda_c +\Delta x/2} \left( 1 - \frac{F_{\lambda, \, \text{line}}}{F_{\lambda, \, \text{cont}}} \right) \, d\lambda,
    \label{eq:equivalent_width}
\end{equation}
where $\lambda_c$ and $\Delta x$ are the central wavelength and the wavelength span of the spectral feature respectively (see Table \ref{EW_bands}), $F_{\lambda,\,\text{line}}$ is the flux density of the spectrum at each target spectral line, and $F_{\lambda,\,\text{cont}}$ is the fitted continuum (equation \ref{cont_eq}) under the corresponding spectral line based on the fit to the continuum bands. Emission lines result in negative EWs by this convention (in accordance with SDSS). For the H$\alpha$ + [\ion{N}{II}] $\lambda \lambda$6548,84 blend, no Gaussian fitting was performed to deblend and estimate the individual contributions of each component. Instead, the combined EW is measured as a single spectral feature. Figure \ref{spectrum_line_ranges} illustrates an example SDSS spectrum, where the spectral lines of interest have been annotated along with the spectral range (see Table \ref{EW_bands}) utilized for calculating the EW.

\end{appendix}
\end{document}